\documentclass[12pt,reqno]{article}
\PassOptionsToPackage{linktocpage}{hyperref}
\usepackage{setspace}
\usepackage{fullpage}
\usepackage{physics}
\usepackage{amsmath}
\usepackage{tikz}
\usepackage{mathdots}
\usepackage{yhmath}
\usepackage{cancel}
\usepackage{color}
\usepackage{siunitx}
\usepackage{array}
\usepackage{multirow}
\usepackage{amssymb}
\usepackage{textcomp,gensymb}
\usepackage{tabularx}
\usepackage{booktabs}
\usetikzlibrary{fadings}
\usepackage{graphicx}
\usepackage{float}
\usepackage{comment}
\usepackage{physics}
\usepackage{amsmath}
\usepackage{tikz}
\usepackage{mathdots}
\usepackage{yhmath}
\usepackage{cancel}
\usepackage{color}
\usepackage{siunitx}
\usepackage{array}
\usepackage{epsfig}
\usepackage{multirow}
\usepackage{amssymb}
\usepackage{gensymb}
\usepackage{tabularx}
\usepackage{booktabs}
\usetikzlibrary{fadings}
\usetikzlibrary{patterns}
\usetikzlibrary{shadows.blur}
\usetikzlibrary{shapes}
\usepackage[numbers,sort&compress]{natbib}
\usepackage{appendix}

\begin{document}
\begin{titlepage}
\hfill \\
\vspace*{15mm}
\begin{center}
{\Large \bf de Sitter Complementarity, TCC, and the Swampland}

\vspace*{15mm}

{\large Alek Bedroya}
\vspace*{8mm}

\textit{Jefferson Physical Laboratory, Harvard University, Cambridge, MA 02138, USA}\\

\vspace*{0.7cm}
%%\maketitle

\end{center}
\begin{abstract}
Motivated by the coincidence of scrambling time in de Sitter space and maximum lifetime given by the \textit{Trans-Planckian Censorship Conjecture} (TCC), we study the relation between the de Sitter complementarity and the Swampland conditions. We study thermalization in de Sitter space from different perspectives and show that TCC implies de Sitter space cannot live long enough to be considered a thermal background. We also revisit $\alpha$-vacua in light of this work and show that TCC imposes multiple initial condition/fine-tuning problems on any conventional inflationary scenario. 

\end{abstract}

\end{titlepage}

\tableofcontents

\section{Introduction}\label{1}

It is remarkably more challenging to construct a de Sitter vacuum in string theory than a flat or an AdS vacuum. In the past few years, several Swampland conditions have been proposed that aim to pinpoint a mutual property among theories in the Landscape that could explain this hurdle in string theory \cite{Obied:2018sgi,Ooguri:2018wrx,Bedroya:2019snp}. However, string theory is not the only place where de Sitter space sets itself apart from flat and AdS spaces. Another notable example is its finite-sized Hilbert space and thermal properties, which are absent from other backgrounds. It is natural to think that all the unique features of de Sitter space should be fundamentally inter-connected. If true, there should be some relation between the Swampland program and the thermodynamic features of de Sitter space.

Interestingly, in the study of \textit{Trans-Planckian Censorship Conjecture} (TCC) in \cite{Bedroya:2019snp}, a strange coincidence strongly suggestive of such a connection was noted. The observation was that the maximum allowed de Sitter space lifetime by TCC matches the scrambling time of de Sitter space! At first, this might seem strange as one time-scale is motivated by string theory, and the other comes from de Sitter complementarity. However, as we discussed earlier, this connection is natural since both contexts study a feature of de Sitter space, which sets it apart from flat and AdS backgrounds.

We investigate this non-trivial coincidence to find a thermodynamic interpretation for TCC. The goal of this paper it to take a small step in bridging the gap between the Swampland program and the extensive literature about thermal aspects of de Sitter space.

The organization of the paper is as follows. In section \ref{2}, we study the consequences of the Swampland conditions for the de Sitter space with a particular focus on TCC. However, before that, we mention some of the motivations for TCC that lend support to those results. In section \ref{3}, we study de Sitter space from the lens of de Sitter complementarity and other perspectives that view de Sitter space as a thermal background. We show that all those ideas point towards the same result that if de Sitter space lives long enough, it would be a thermal background with a thermalization time of $\sim \frac{1}{H}\ln(\frac{1}{H})$. In section \ref{4}, we put the Swampland picture next to the other pictures to arrive at a thermodynamic interpretation of TCC. We argue that TCC, in its essence, tells us that de Sitter space is not stable enough to be viewed as a thermal background. We elaborate on the physical meaning of this interpretation and support it by other Swampland conditions. Finally, we show in section \ref{5} that TCC imposes a severe initial condition problem on inflation.

\section{Swampland picture}\label{2}

This section explores the picture that Swampland conditions, especially TCC, offer of de Sitter space. We begin by reviewing TCC and some motivations for it that lend support to its implications. We then study the implications of Swampland conditions for different possible realizations of de Sitter space.  
 
\subsection{TCC and its motivations}

\begin{figure}[H]
    \centering

\tikzset{every picture/.style={line width=0.75pt}} %set default line width to 0.75pt        

\begin{tikzpicture}[x=0.75pt,y=0.75pt,yscale=-1,xscale=1]

%Shape: Rectangle [id:dp3638945827656501] 
\draw  [draw opacity=0][fill={rgb, 255:red, 255; green, 0; blue, 0 }  ,fill opacity=0.57 ][line width=1.5]  (197,61.71) -- (267,61.71) -- (267,250.71) -- (197,250.71) -- cycle ;
%Shape: Rectangle [id:dp03240086923721197] 
\draw  [draw opacity=0][fill={rgb, 255:red, 72; green, 150; blue, 241 }  ,fill opacity=0.69 ][line width=1.5]  (369,61.71) -- (439,61.71) -- (439,250.71) -- (369,250.71) -- cycle ;
%Shape: Axis 2D [id:dp7114600356243816] 
\draw  (192,250.71) -- (494,250.71)(197,39.17) -- (197,268.17) (487,245.71) -- (494,250.71) -- (487,255.71) (192,46.17) -- (197,39.17) -- (202,46.17)  ;
%Straight Lines [id:da43080407651163566] 
\draw [color={rgb, 255:red, 155; green, 155; blue, 155 }  ,draw opacity=1 ] [dash pattern={on 4.5pt off 4.5pt}]  (197,61.71) -- (439,61.71) ;
%Curve Lines [id:da5894566021834569] 
\draw    (255.5,249.71) .. controls (263.5,204.71) and (317.5,85.71) .. (369,61.71) ;
%Curve Lines [id:da698755688167481] 
\draw    (202.5,250.71) .. controls (203.5,223.71) and (226.5,108.71) .. (267,61.71) ;
%Straight Lines [id:da375300134937401] 
\draw    (269.5,53.71) -- (367.5,53.71) ;
\draw [shift={(369.5,53.71)}, rotate = 180] [color={rgb, 255:red, 0; green, 0; blue, 0 }  ][line width=0.75]    (10.93,-3.29) .. controls (6.95,-1.4) and (3.31,-0.3) .. (0,0) .. controls (3.31,0.3) and (6.95,1.4) .. (10.93,3.29)   ;
\draw [shift={(267.5,53.71)}, rotate = 0] [color={rgb, 255:red, 0; green, 0; blue, 0 }  ][line width=0.75]    (10.93,-3.29) .. controls (6.95,-1.4) and (3.31,-0.3) .. (0,0) .. controls (3.31,0.3) and (6.95,1.4) .. (10.93,3.29)   ;

% Text Node
\draw (172,51.11) node [anchor=north west][inner sep=0.75pt]    {$t_{f}$};
% Text Node
\draw (172,242.11) node [anchor=north west][inner sep=0.75pt]    {$t_{i}$};
% Text Node
\draw (192,17.11) node [anchor=north west][inner sep=0.75pt]    {$t$};
% Text Node
\draw (501,240.11) node [anchor=north west][inner sep=0.75pt]    {$\lambda $};
% Text Node
\draw (260,256.11) node [anchor=north west][inner sep=0.75pt]    {$l_{p}$};
% Text Node
\draw (361,256.4) node [anchor=north west][inner sep=0.75pt]    {$\frac{1}{H}$};
% Text Node
\draw (250,24.71) node [anchor=north west][inner sep=0.75pt]   [align=left] {{\fontfamily{ptm}\selectfont {\scriptsize Dynamical degrees of freedom}}};

\end{tikzpicture}

        \caption{The curved lines denote the expansion of wave-lengths of two comoving modes. These two modes correspond to the greatest, and smallest dynamical modes at time $t_f$. As shown in the figure, if TCC is violated over some time interval $[t_i,t_f]$, no comoving mode would stay dynamical throughout that time window.}
    \label{fig:my_label7}
\end{figure}
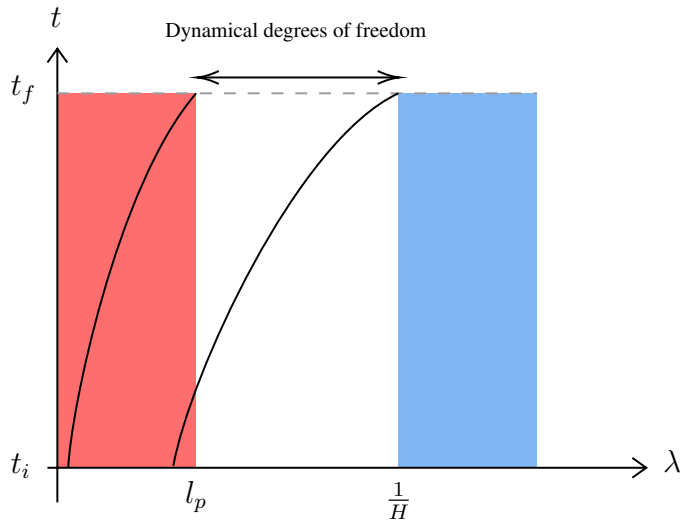

\textit{Trans-Planckian censorship conjecture} (TCC) postulates that in a consistent quantum theory of gravity, an expansionary universe in which Planckian modes exit the Hubble horizon cannot be realized \cite{Bedroya:2019snp}. What is special about the Hubble radius is that when a mode exits the Hubble horizon, it becomes non-dynamical and freezes out \cite{Martin:2000xs,Brandenberger:2000wr,Brandenberger:2012aj}. Moreover, super-Hubble modes undergo decoherence which makes them equivalent to stochastic classical perturbations \cite{Polarski:1995jg} and the modes will remain classical even if/when they re-enter the Hubble horizon. As shown in figure \ref{fig:my_label7}, a violation of TCC would lead to the classicalization of all dynamical quantum fluctuations $H<k<l_{P}^{-1}$.

 Following we review some of the arguments in favor of TCC.

\subsubsection{Central node in the Swampland web}

The TCC implies some versions of several Swampland conditions. These versions have shown to be true in known controllable string theory constructions. These non-trivial consistencies strongly support TCC. The connections between TCC and other Swampland conditions suggest that TCC is a central node in the web of Swampland conditions. One that brings many of them together and provides a simple physical explanation for them. Following, we review some of the non-trivial implications of TCC that resemble other Swampland conditions and their consistency with string theory.
\vspace{20pt}

\begin{figure}
    \centering

\tikzset{every picture/.style={line width=0.75pt}} %set default line width to 0.75pt        

\begin{tikzpicture}[x=0.75pt,y=0.75pt,yscale=-1,xscale=1]

%Rounded Rect [id:dp7325789943629673] 
\draw  [fill={rgb, 255:red, 81; green, 203; blue, 72 }  ,fill opacity=0.69 ] (288,135.71) .. controls (288,131.3) and (291.58,127.71) .. (296,127.71) -- (350,127.71) .. controls (354.42,127.71) and (358,131.3) .. (358,135.71) -- (358,159.71) .. controls (358,164.13) and (354.42,167.71) .. (350,167.71) -- (296,167.71) .. controls (291.58,167.71) and (288,164.13) .. (288,159.71) -- cycle ;
%Up Arrow [id:dp5876471237048315] 
\draw  [fill={rgb, 255:red, 33; green, 180; blue, 151 }  ,fill opacity=0.58 ] (348.45,77.82) -- (369.16,64.53) -- (369.6,89.14) -- (364.31,86.31) -- (349.1,114.73) -- (338.53,109.07) -- (353.74,80.65) -- cycle ;
%Rounded Rect [id:dp5872688110171069] 
\draw  [fill={rgb, 255:red, 137; green, 166; blue, 239 }  ,fill opacity=1 ] (361,25.71) .. controls (361,21.3) and (364.58,17.71) .. (369,17.71) -- (495.5,17.71) .. controls (499.92,17.71) and (503.5,21.3) .. (503.5,25.71) -- (503.5,49.71) .. controls (503.5,54.13) and (499.92,57.71) .. (495.5,57.71) -- (369,57.71) .. controls (364.58,57.71) and (361,54.13) .. (361,49.71) -- cycle ;

%Rounded Rect [id:dp48680395706343305] 
\draw  [fill={rgb, 255:red, 137; green, 166; blue, 239 }  ,fill opacity=1 ] (413,224.71) .. controls (413,220.3) and (416.58,216.71) .. (421,216.71) -- (573.5,216.71) .. controls (577.92,216.71) and (581.5,220.3) .. (581.5,224.71) -- (581.5,248.71) .. controls (581.5,253.13) and (577.92,256.71) .. (573.5,256.71) -- (421,256.71) .. controls (416.58,256.71) and (413,253.13) .. (413,248.71) -- cycle ;

%Rounded Rect [id:dp4563356004386141] 
\draw  [fill={rgb, 255:red, 137; green, 166; blue, 239 }  ,fill opacity=1 ] (201,247.71) .. controls (201,243.3) and (204.58,239.71) .. (209,239.71) -- (390.5,239.71) .. controls (394.92,239.71) and (398.5,243.3) .. (398.5,247.71) -- (398.5,271.71) .. controls (398.5,276.13) and (394.92,279.71) .. (390.5,279.71) -- (209,279.71) .. controls (204.58,279.71) and (201,276.13) .. (201,271.71) -- cycle ;

%Rounded Rect [id:dp9566420996870457] 
\draw  [fill={rgb, 255:red, 137; green, 166; blue, 239 }  ,fill opacity=1 ] (6,151.71) .. controls (6,147.3) and (9.58,143.71) .. (14,143.71) -- (200.5,143.71) .. controls (204.92,143.71) and (208.5,147.3) .. (208.5,151.71) -- (208.5,175.71) .. controls (208.5,180.13) and (204.92,183.71) .. (200.5,183.71) -- (14,183.71) .. controls (9.58,183.71) and (6,180.13) .. (6,175.71) -- cycle ;

%Rounded Rect [id:dp961569224215985] 
\draw  [fill={rgb, 255:red, 137; green, 166; blue, 239 }  ,fill opacity=1 ] (447,114.71) .. controls (447,110.3) and (450.58,106.71) .. (455,106.71) -- (613.5,106.71) .. controls (617.92,106.71) and (621.5,110.3) .. (621.5,114.71) -- (621.5,138.71) .. controls (621.5,143.13) and (617.92,146.71) .. (613.5,146.71) -- (455,146.71) .. controls (450.58,146.71) and (447,143.13) .. (447,138.71) -- cycle ;

%Rounded Rect [id:dp6827849450903765] 
\draw  [fill={rgb, 255:red, 137; green, 166; blue, 239 }  ,fill opacity=1 ] (61,28.71) .. controls (61,24.3) and (64.58,20.71) .. (69,20.71) -- (307.5,20.71) .. controls (311.92,20.71) and (315.5,24.3) .. (315.5,28.71) -- (315.5,52.71) .. controls (315.5,57.13) and (311.92,60.71) .. (307.5,60.71) -- (69,60.71) .. controls (64.58,60.71) and (61,57.13) .. (61,52.71) -- cycle ;

%Up Arrow [id:dp19399843337253686] 
\draw  [fill={rgb, 255:red, 33; green, 180; blue, 151 }  ,fill opacity=0.52 ] (405.72,124.14) -- (432.9,130.28) -- (411.02,147.54) -- (409.69,141.69) -- (372.89,150.03) -- (370.24,138.33) -- (407.04,129.99) -- cycle ;
%Up Arrow [id:dp47190867839600137] 
\draw  [fill={rgb, 255:red, 33; green, 180; blue, 151 }  ,fill opacity=0.5 ] (390.81,192.55) -- (397.54,216.83) -- (373.53,209.19) -- (377.85,205.03) -- (354.79,181.1) -- (363.43,172.78) -- (386.49,196.71) -- cycle ;
%Up Arrow [id:dp7560585182723822] 
\draw  [fill={rgb, 255:red, 33; green, 180; blue, 151 }  ,fill opacity=0.56 ] (325.44,211.5) -- (310.54,230.77) -- (301.69,208.07) -- (307.63,208.93) -- (312.16,177.45) -- (324.03,179.17) -- (319.5,210.64) -- cycle ;
%Up Arrow [id:dp6046034263074067] 
\draw  [fill={rgb, 255:red, 33; green, 180; blue, 151 }  ,fill opacity=0.51 ] (240.77,172.06) -- (212.56,164.25) -- (237.13,148.35) -- (238.04,154.27) -- (277.62,148.2) -- (279.44,160.05) -- (239.86,166.13) -- cycle ;
%Up Arrow [id:dp7982543218241742] 
\draw  [fill={rgb, 255:red, 33; green, 180; blue, 151 }  ,fill opacity=0.54 ] (254.83,94.17) -- (252.52,66.82) -- (275.45,81.92) -- (270.3,84.98) -- (289.22,116.83) -- (278.91,122.96) -- (259.98,91.11) -- cycle ;

% Text Node
\draw (387.94,28.71) node [anchor=north west][inner sep=0.75pt]   [align=left] {dS Conjecture};
% Text Node
\draw (23.64,154.71) node [anchor=north west][inner sep=0.75pt]   [align=left] {Weak Gravity Conjecture};
% Text Node
\draw (457.82,117.71) node [anchor=north west][inner sep=0.75pt]   [align=left] {Refined dS Conjecture};
% Text Node
\draw (215.03,250.71) node [anchor=north west][inner sep=0.75pt]   [align=left] {AdS Distance Conjecture};
% Text Node
\draw (427.86,227.71) node [anchor=north west][inner sep=0.75pt]   [align=left] {Distance Conjecture};
% Text Node
\draw (84.25,31.71) node [anchor=north west][inner sep=0.75pt]   [align=left] {No eternal Inflation Conjecture};
% Text Node
\draw (307,138.71) node [anchor=north west][inner sep=0.75pt]   [align=left] {{\fontfamily{ptm}\selectfont \textbf{TCC}}};

\end{tikzpicture}

    \caption{TCC non-trivially implies some version of a number of Swampland conditions. This is remarkable given how simple the physical idea behind TCC is.}
    \label{fig:my_label6}
\end{figure}
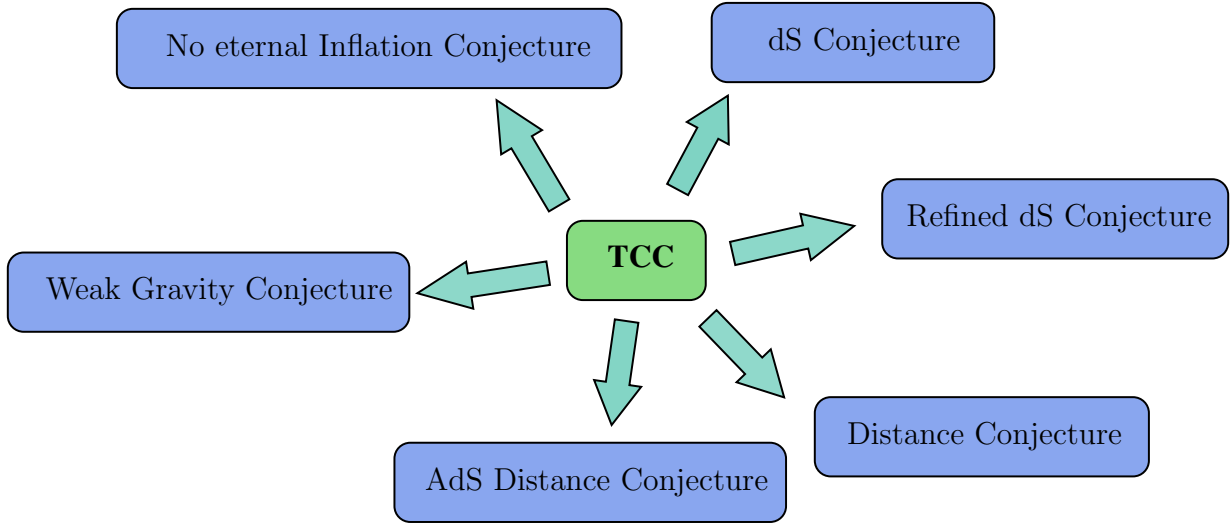

\bf The de Sitter Conjecture\normalsize\normalfont: One of the most notable implication of TCC is that in d-non-compact dimensions, $\frac{|\nabla V|}{V}\geq \frac{2}{\sqrt{(d-1)(d-2)}}$ in the asymptotics of the field space. This is similar to the de Sitter conjecture \cite{Obied:2018sgi}, but it provides a definite lower bound which is remarkably satisfied in multitudes of string theory constructions \cite{Bedroya:2019snp,Andriot:2020lea}. 
\vspace{20pt}

\bf Distance conjecture \normalsize\normalfont: As proposed in \cite{Andriot:2020lea}, TCC suggests a definite lower bound $\frac{1}{\sqrt{6}}$ for the order one constant $\lambda$ in the distance conjecture in 4d \cite{Ooguri:2006in}. This bound could also be motivated by the following heuristic argument. Suppose $\phi_1$ and $\phi_2$ are canonically normalized fields where $\phi_2$ is a scalar field in a tower of light states emerging as $\phi_1\rightarrow\infty$. As $\phi_1$ goes to infinity, mass of $\phi_2$ exponentially decays. After some point, $\phi_2$ becomes light enough to be added to the spectrum of the low energy field theory. In this regime, the potential depends on both $\phi_1$ and $\phi_2$ and the mass of $\phi_2$ is roughly given by $m\sim \sqrt{{\partial_{\phi_2}^2V}}$. In conventional string theory constructions, the potential decays exponentially in asymptotic directions. Suppose the potential behaves as $V\sim f(\phi_2)\exp(-g(\phi_2)\phi_1)$ for large $\phi_1$, the mass scale $m$ would decay like $\sim \exp(-\frac{g(\phi_2)}{2}\phi_1)$ as $\phi_1$ goes to infinity. From TCC, we know that $g(\phi_2)\geq\frac{2}{\sqrt{(d-1)(d-2)}}$ \cite{Bedroya:2019snp} which leads to 
\begin{align}\label{TCCdis}
    \lambda\geq \lambda_{TCC}=\frac{1}{\sqrt{(d-1)(d-2)}}.
\end{align}
In \cite{Grimm:2018ohb}, authors found a lower bound for $\lambda$
\begin{align}
    &\lambda\geq\sqrt\frac{1}{10-d}~~~\frac{d}{2}=even\nonumber\\
    &\lambda\geq\sqrt\frac{2}{10-d}~~~\frac{d}{2}=odd,
\end{align}
for single modulus limit in various Calabi-Yau compactifications. Remarkably, this bound is stronger that the TCC-motivated bound \eqref{TCCdis} for every $d\geq4$. The TCC-motivated bound has been checked more generally in a variety of 4d string theory constructions in \cite{Gendler:2020dfp,Andriot:2020lea}. 
\vspace{20pt}

\bf Refined de Sitter conjecture\normalsize\normalfont: The refined de Sitter conjecture states that there is a universal $\mathcal{O}(1)$ lower bound for $|\Delta V|/V$ at local maxima \cite{Ooguri:2018wrx,Garg:2018reu}. Interestingly, TCC implies a logarithmically corrected version of this condition for local maxima \cite{Bedroya:2019snp}. The modified condition roughly is $\frac{|\Delta V|}{V}>\frac{16}{(d-1)(d-2)\ln(V)^2}$.
 \vspace{20pt}
 
\bf Weak gravity conjecture (WGC) and generalized distance conjecture\normalsize\normalfont: Consider a flux generated potential with a charged co-dimension 1 brane. The brane serves as the domain wall for tunneling between neighbouring vacua. TCC implies both the WGC and the generalized distance conjecture for the brane in all dimensions \cite{Bedroya:2020rac}. 
\vspace{20pt}

\bf No eternal inflation\normalsize\normalfont: Suppose tunnellings between neighbouring vacua are non-drastic enough that a monotonic quintessence potential could effectively describe the universe's evolution. In that case, one can show TCC marginally forbids eternal inflation in any dimension \cite{Bedroya:2020rac}. 

Figure \ref{fig:my_label6} shows all the Swampland conjectures that, in some form, are implied by TCC.

\subsubsection{Coincidence problem}

An immediate consequence of $TCC$ is that the age of the dark energy dominated epoch $T_\Lambda$ must be less than $\frac{1}{H}\ln(\frac{1}{H})$ which is true in our universe. This consistency already provides a simple, yet non-trivial, experimental test for TCC. Perhaps the more interesting fact is that our universe only marginally satisfies this inequality thanks to the logarithmic term $\ln(\frac{1}{H})$. This "accident" is precisely the coincidence problem in cosmology. If the cosmological constant is a constant, i.e. the universe is stuck in a local minimum of the potential, there is no apriori reason for $T_\Lambda\sim\frac{1}{H}$. Statistically speaking, for most of the lifetime of a metastable universe, its age is of the order of its lifetime, and TCC relates that lifetime to the Hubble time. In fact, according to TCC, no matter the universe is in metastable equilibrium, or it is rolling, the coincidence of $T\sim\frac{1}{H}$ is anticipated, which is another non-trivial consistency with observation. 

\subsubsection{TCC as gravitational renormalizability}

We argue that TCC could be viewed as a natural modification of the renormalizability condition for gravitational theories where the conventional notion of renormalizability does not apply. To see this, let us review what renormalizability means in field theories.

In low energy field theories, the high energy modes ($k>E$) are assumed to be in their ground states. The ground states are non-classical superpositions of classical waves, which could also be understood as quantum fluctuations. If classical perturbations could grow into or impact the time evolution of low-energy classical modes, the quantum superposition of UV modes would extend into the low-energy regime. This would make it impossible to have a well-defined classical low-energy theory. In the quantum language, this is a consequence of unitarity in the subspace where all the UV modes are in their corresponding ground states. Such scale separation is the meaning of renormalizabilty. Therefore, in flat space, renormalizability naively implies that there exists a momentum cutoff $\Lambda$ such that classical high momentum perturbations with $k> \Lambda$ do not influence the dynamics of low energy modes with $K<\Lambda$. However, this naive notion of scale separation does not apply to GR since any expansion of the universe stretches some modes with $k>\Lambda$ into modes with $k<\Lambda$. The above naive argument is a simple way of understanding why GR is non-renormalizable. 

Similar to how quantum field theory's consistency imposes renormalizability, it is natural to expect a UV-complete quantum theory of gravity must satisfy some renormalizability-like condition. As we saw earlier, the naive scale separation (renormalizability) does not hold in gravitational theories. The simplest relaxation to a scale separation between classical modes with $k>\Lambda$ and $k<\Lambda$, is to postulate that there are \bf two energy scales $\Lambda_{UV}\gg\Lambda_{IR}$ \normalfont such that the deep UV modes $k>\Lambda_{UV}$ do not stretch into deep IR modes $k<\Lambda_{IR}$. There are natural candidates for these energy scales in any de Sitter background; the Planck scale and the Hubble scale. Using these scales, our candidate for gravitational renormalizability takes the following form.
\vspace{10pt}

\bf Modes with $k> \frac{1}{l_P}$ or equivalently $\lambda<l_P$ cannot evolve into modes with $k<H$ or equivalently $\lambda>\frac{1}{H}$. \normalfont
\vspace{10pt}

In other words, sub-Planckian modes cannot exit the Hubble horizon, which is precisely the statement of TCC. In a sense, TCC is a natural gravitational analogue of the renormalizability condition in field theory.

\subsubsection{Initial condition problem for inflation}

We want to take this opportunity to clarify a possible confusion that a particular initial condition problem for inflation, has been a motivation for TCC. A violation of TCC poses two apparent initial condition problems for inflation. We briefly review each of these problems, the resolutions proposed in the literature, and how they relate to TCC.

\bf First initial condition problem:\normalfont

If some fluctuations, e.g. Hubble sized CMB fluctuations, trace back to trans-Planckian fluctuations, it seems part of the needed initial condition is inaccessible to the field theory. This raises a practical question. What initial state should we consider for those modes as they become sub-Planckian and enter the range of field theory? 

\bf Resolution: \normalfont

Requiring the vacuum to be like the Minkowski vacuum at short distances fixes the de Sitter vacuum at Planckian momenta \cite{Kaloper:2002cs}. Essentially, the equivalence principle naturally screens the trans-Planckian physics from sub-Planckian observers. This argument resolves the problem for as long as de Sitter space is stable as a semi-classical background. This "problem" has not been a motivation for TCC, and the mentioned resolution is not affected by TCC.

\bf Second initial condition problem:\normalfont

In contrary to the Minkowski space, the de Sitter space does not have a unique vacuum. There is a family of vacua called $\alpha$-vacua that all are invariant under symmetries of de Sitter space. However, the only $\alpha$-vacuum that would give scale-invariant CMB fluctuations is the Bunch-Davies (BD) vacuum. Why did the universe choose this particular vacuum state in the inflationary era?

\bf Resolution: \normalfont

The authors in \cite{Kaloper:2002cs} argued that any deviation from the BD vacuum eventually exits the Hubble horizon and disappears. Therefore, inflation automatically sets the universe in the BD vacuum. We revisit the de Sitter vacua and this argument in section \ref{3}, and we contrast it with TCC in section \ref{4}. We show that the argument in \cite{Kaloper:2002cs} makes an assumption that is fundamentally inconsistent with TCC. In other words, a violation of TCC is baked in the argument. Assuming TCC is correct, this argument no longer works. We will come back to this initial condition problem later in section \ref{5}.

\subsection{de Sitter space and the Swampland}

Swampland conditions suggest that the de Sitter space is in tension with a UV-complete theory of gravity. The de Sitter conjecture forbids the de Sitter space all-together, but TCC allows it as long as it is sufficiently short-lived. TCC requires the de Sitter space to undergo some significant transformation by $\tau_{TCC}=\frac{1}{H}\ln(\frac{1}{H})$ so that the Planckain fluctuations do not exit the Hubble horizon. This transformation could be a significant drop in cosmological constant, tunnelling to a different vacuum, quantum breaking of spacetime, or the effective field theory's breakdown. Following, we study each of these possibilities separately. 

Suppose the cosmological constant continuously discharges (quintessence), TCC tells us that a major part of $\Lambda$ have to be discharged by $\tau_{TCC}$. This implies that the quintessence potential could not be too flat. For monotonic potentials, this leads to $\frac{|V'|}{V}\gtrsim \frac{2}{\sqrt{(d-1)(d-2)}}$ in asymptotics of the field space and $\frac{|V'|}{V}\gtrsim \mathcal{O}(\ln(V)^{-2})$ in the interior of the field space \cite{Bedroya:2019snp}. For local maxima on the other hand, TCC roughly implies $\frac{|V''|}{V}\gtrsim \frac{16}{(d-1)(d-2)\ln(V)^2}$ \cite{Bedroya:2019snp}.

Suppose the universe undergoes tunnelling, i.e. a bubble of a more stable vacuum forms and expands until it takes over the Hubble patch. TCC implies the metastable vacuum's lifetime must be less than $\tau_{TCC}$. This has important implications for the domain wall of the bubble. In particular, the domain wall must satisfy both WGC and the generalized distance conjecture \cite{Bedroya:2020rac}. 

Another way to avoid violating TCC is that the QFT in curved background description breaks down. This could happen in two ways; 1) Break down of the classical background, often referred to as quantum breaking, 2) Break down of the EFT due to the emergence of new light states in the theory. Whichever happens first, TCC tells us that it must take place by $\tau_{TCC}$. Authors in \cite{Dvali:2013eja,Dvali:2014gua,Dvali:2017eba} argued that the quantum breaking time for de Sitter space is $\sim H^{-3}$, which is greater than $\tau_{TCC}$. However, the de Sitter and the distance conjectures show that the EFT breaks down by $\tau_{TCC}$ in the asymptotics of the field space \cite{Seo:2019wsh}.

All in all, we expect the de Sitter space to undergo some significant physical transformation before $\tau_{TCC}$. No matter which scenario happens, we come across the same timescale $\tau_{TCC}$, after which the initial de Sitter description should no longer work.

\section{Complementarity picture}\label{3}

In this section, we study the de Sitter space from the complementarity perspective. First, we review black hole complementarity from which most of the ideas for de Sitter complementarity have originated. As we will see, black holes do not share some of the strange features of de Sitter space, which makes black hole complementarity simpler and easier to understand than its de Sitter counterpart. 

\subsection{Review of black hole complementarity}\label{rbhc}

In nutshell, the black hole complementarity is a partial resolution to a tension between well-established physical principles. We briefly review the argument in \cite{Susskind:1993if} that leads to the idea of complementarity. 

We assume the following postulates\footnote{Our list of postulates is slightly different from that of \cite{Susskind:1993if}, but the following argument is almost identical.}:

1) \textit{Unitary semi-classical QFT in curved spacetime}: We treat gravity classically and other fields quantum mechanically. We assume every spacelike Cauchy surface has an associated Hilbert space for the quantum fields living on it. The time evolution between any two such Cauchy surfaces is given by a time-dependent unitary operator dependent on the gravitational background.

2) \textit{Equivalence principle}: Every free-falling observer must be unable to distinguish the spacetime from Minkowski space through performing local experiments. 

3) \textit{No remnant}: We assume the black hole will completely evaporate at a finite time. 

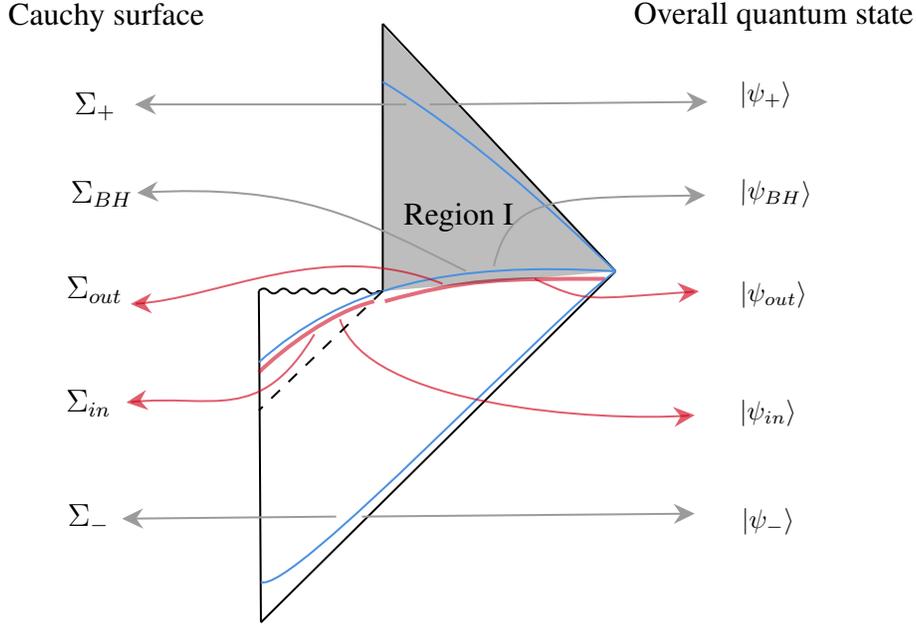
\begin{figure}
    \centering

\tikzset{every picture/.style={line width=0.75pt}} %set default line width to 0.75pt        

\begin{tikzpicture}[x=0.75pt,y=0.75pt,yscale=-1,xscale=1]
%uncomment if require: \path (0,327); %set diagram left start at 0, and has height of 327

%Curve Lines [id:da7434426379440247] 
\draw [color={rgb, 255:red, 208; green, 2; blue, 27 }  ,draw opacity=0.58 ][line width=1.5]    (283.06,156.8) .. controls (324.02,145.38) and (341.26,145.38) .. (394.07,145.45) ;
%Shape: Triangle [id:dp5211557748564022] 
\draw  [color={rgb, 255:red, 0; green, 0; blue, 0 }  ,draw opacity=0 ][fill={rgb, 255:red, 0; green, 0; blue, 0 }  ,fill opacity=0.26 ] (281.98,151.68) -- (281.46,17.37) -- (399.45,141.63) -- cycle ;
%Straight Lines [id:da309402010683431] 
\draw    (219.47,151.57) -- (219.7,179.86) -- (220.55,318.86) ;
%Straight Lines [id:da857067298061249] 
\draw    (219.47,151.57) .. controls (221.14,149.91) and (222.81,149.91) .. (224.47,151.58) .. controls (226.14,153.25) and (227.8,153.25) .. (229.47,151.59) .. controls (231.14,149.93) and (232.8,149.93) .. (234.47,151.6) .. controls (236.14,153.27) and (237.8,153.27) .. (239.47,151.61) .. controls (241.14,149.95) and (242.8,149.95) .. (244.47,151.62) .. controls (246.14,153.29) and (247.8,153.29) .. (249.47,151.63) .. controls (251.14,149.97) and (252.8,149.97) .. (254.47,151.64) .. controls (256.14,153.31) and (257.8,153.31) .. (259.47,151.65) .. controls (261.14,149.99) and (262.8,149.99) .. (264.47,151.66) .. controls (266.14,153.33) and (267.8,153.33) .. (269.47,151.67) .. controls (271.14,150.01) and (272.8,150.01) .. (274.47,151.68) .. controls (276.14,153.35) and (277.8,153.35) .. (279.47,151.69) -- (281.98,151.7) -- (281.98,151.7) ;
%Straight Lines [id:da19827822567073983] 
\draw    (281.98,16.67) -- (281.98,151.7) ;
%Straight Lines [id:da8616412981414496] 
\draw    (399.46,141.63) -- (220.55,318.86) ;
%Straight Lines [id:da9920212563560673] 
\draw    (281.98,16.67) -- (399.46,141.63) ;
%Curve Lines [id:da7930373055494102] 
\draw [color={rgb, 255:red, 74; green, 144; blue, 226 }  ,draw opacity=1 ]   (220.5,298.43) .. controls (235.5,305.43) and (368.5,163.43) .. (399.46,141.63) ;
%Curve Lines [id:da42244813330929] 
\draw [color={rgb, 255:red, 74; green, 144; blue, 226 }  ,draw opacity=1 ]   (219.47,187.53) .. controls (262.58,149.28) and (314.32,136.53) .. (399.46,141.63) ;
%Curve Lines [id:da2691710667373921] 
\draw [color={rgb, 255:red, 74; green, 144; blue, 226 }  ,draw opacity=1 ]   (281.98,46) .. controls (336.95,79.15) and (390.84,133.98) .. (399.46,141.63) ;
%Straight Lines [id:da46350292602336207] 
\draw [color={rgb, 255:red, 155; green, 155; blue, 155 }  ,draw opacity=1 ]   (305.69,57.4) -- (442.81,56.15) ;
\draw [shift={(445.81,56.12)}, rotate = 539.48] [fill={rgb, 255:red, 155; green, 155; blue, 155 }  ,fill opacity=1 ][line width=0.08]  [draw opacity=0] (10.72,-5.15) -- (0,0) -- (10.72,5.15) -- (7.12,0) -- cycle    ;
%Straight Lines [id:da8950436533033512] 
\draw [color={rgb, 255:red, 155; green, 155; blue, 155 }  ,draw opacity=1 ]   (293.84,57.4) -- (159.96,57.4) ;
\draw [shift={(156.96,57.4)}, rotate = 360] [fill={rgb, 255:red, 155; green, 155; blue, 155 }  ,fill opacity=1 ][line width=0.08]  [draw opacity=0] (10.72,-5.15) -- (0,0) -- (10.72,5.15) -- (7.12,0) -- cycle    ;
%Curve Lines [id:da22272428742576866] 
\draw [color={rgb, 255:red, 208; green, 2; blue, 27 }  ,draw opacity=0.58 ][line width=1.5]    (219.47,192.63) .. controls (231.33,181.08) and (256.12,161.95) .. (277.67,156.8) ;
%Curve Lines [id:da7571935547156674] 
\draw [color={rgb, 255:red, 208; green, 2; blue, 27 }  ,draw opacity=0.64 ]   (312.16,147.93) .. controls (253.01,124.19) and (199.04,155.55) .. (157.34,158.01) ;
\draw [shift={(154.81,158.13)}, rotate = 358.26] [fill={rgb, 255:red, 208; green, 2; blue, 27 }  ,fill opacity=0.64 ][line width=0.08]  [draw opacity=0] (10.72,-5.15) -- (0,0) -- (10.72,5.15) -- (7.12,0) -- cycle    ;
%Curve Lines [id:da681523579469467] 
\draw [color={rgb, 255:red, 208; green, 2; blue, 27 }  ,draw opacity=0.64 ]   (247.49,173.43) .. controls (220.03,219.66) and (195.68,203.44) .. (155.15,207.58) ;
\draw [shift={(152.65,207.86)}, rotate = 353.08000000000004] [fill={rgb, 255:red, 208; green, 2; blue, 27 }  ,fill opacity=0.64 ][line width=0.08]  [draw opacity=0] (10.72,-5.15) -- (0,0) -- (10.72,5.15) -- (7.12,0) -- cycle    ;
%Curve Lines [id:da9325808012514143] 
\draw [color={rgb, 255:red, 155; green, 155; blue, 155 }  ,draw opacity=1 ]   (324.02,141.55) .. controls (270.67,115.04) and (240.56,97.28) .. (160.48,101.88) ;
\draw [shift={(158.04,102.03)}, rotate = 356.44] [fill={rgb, 255:red, 155; green, 155; blue, 155 }  ,fill opacity=1 ][line width=0.08]  [draw opacity=0] (10.72,-5.15) -- (0,0) -- (10.72,5.15) -- (7.12,0) -- cycle    ;
%Straight Lines [id:da4113113732682836] 
\draw [color={rgb, 255:red, 155; green, 155; blue, 155 }  ,draw opacity=1 ]   (258.5,265.43) -- (153.49,266.48) ;
\draw [shift={(150.49,266.51)}, rotate = 359.43] [fill={rgb, 255:red, 155; green, 155; blue, 155 }  ,fill opacity=1 ][line width=0.08]  [draw opacity=0] (10.72,-5.15) -- (0,0) -- (10.72,5.15) -- (7.12,0) -- cycle    ;
%Curve Lines [id:da3457735957500161] 
\draw [color={rgb, 255:red, 155; green, 155; blue, 155 }  ,draw opacity=1 ]   (338.03,139) .. controls (346.56,102.4) and (369.89,104.53) .. (442.51,103.34) ;
\draw [shift={(444.73,103.3)}, rotate = 539.02] [fill={rgb, 255:red, 155; green, 155; blue, 155 }  ,fill opacity=1 ][line width=0.08]  [draw opacity=0] (10.72,-5.15) -- (0,0) -- (10.72,5.15) -- (7.12,0) -- cycle    ;
%Curve Lines [id:da40452203745031623] 
\draw [color={rgb, 255:red, 208; green, 2; blue, 27 }  ,draw opacity=0.64 ]   (358.51,145.38) .. controls (385.05,157.94) and (384.4,155.65) .. (438.96,151.92) ;
\draw [shift={(441.49,151.75)}, rotate = 536.1700000000001] [fill={rgb, 255:red, 208; green, 2; blue, 27 }  ,fill opacity=0.64 ][line width=0.08]  [draw opacity=0] (10.72,-5.15) -- (0,0) -- (10.72,5.15) -- (7.12,0) -- cycle    ;
%Curve Lines [id:da3023568188502228] 
\draw [color={rgb, 255:red, 208; green, 2; blue, 27 }  ,draw opacity=0.64 ]   (260.43,165.78) .. controls (273.1,209.51) and (390.32,217.73) .. (436.6,214.45) ;
\draw [shift={(439.34,214.23)}, rotate = 535.05] [fill={rgb, 255:red, 208; green, 2; blue, 27 }  ,fill opacity=0.64 ][line width=0.08]  [draw opacity=0] (10.72,-5.15) -- (0,0) -- (10.72,5.15) -- (7.12,0) -- cycle    ;
%Straight Lines [id:da03968349631989487] 
\draw [color={rgb, 255:red, 155; green, 155; blue, 155 }  ,draw opacity=1 ]   (271.5,265.43) -- (436.34,263.98) ;
\draw [shift={(439.34,263.96)}, rotate = 539.5] [fill={rgb, 255:red, 155; green, 155; blue, 155 }  ,fill opacity=1 ][line width=0.08]  [draw opacity=0] (10.72,-5.15) -- (0,0) -- (10.72,5.15) -- (7.12,0) -- cycle    ;
%Straight Lines [id:da9774618561120874] 
\draw  [dash pattern={on 4.5pt off 4.5pt}]  (281.97,151.68) -- (219.49,211.86) ;

% Text Node
\draw (460.48,44.36) node [anchor=north west][inner sep=0.75pt]  [font=\footnotesize]  {$\ket{\psi _{+}}$};
% Text Node
\draw (461.52,259.84) node [anchor=north west][inner sep=0.75pt]  [font=\footnotesize]  {$\ket{\psi _{-}}$};
% Text Node
\draw (91.7,4.82) node [anchor=north west][inner sep=0.75pt]   [align=left] {{\fontfamily{ptm}\selectfont Cauchy surface}};
% Text Node
\draw (407.13,4.14) node [anchor=north west][inner sep=0.75pt]   [align=left] {{\fontfamily{ptm}\selectfont Overall quantum state}};
% Text Node
\draw (460.9,93.65) node [anchor=north west][inner sep=0.75pt]  [font=\footnotesize,rotate=-1.31]  {$\ket{\psi _{BH}}$};
% Text Node
\draw (124.98,49.76) node [anchor=north west][inner sep=0.75pt]    {$\Sigma _{+}$};
% Text Node
\draw (123.14,94.38) node [anchor=north west][inner sep=0.75pt]    {$\Sigma _{BH}$};
% Text Node
\draw (121.71,257.59) node [anchor=north west][inner sep=0.75pt]    {$\Sigma _{-}$};
% Text Node
\draw (120.75,200.21) node [anchor=north west][inner sep=0.75pt]    {$\Sigma _{in}$};
% Text Node
\draw (120.98,142.83) node [anchor=north west][inner sep=0.75pt]    {$\Sigma _{out}$};
% Text Node
\draw (460.72,145.09) node [anchor=north west][inner sep=0.75pt]  [font=\footnotesize]  {$\ket{\psi _{out}}$};
% Text Node
\draw (460.56,205.02) node [anchor=north west][inner sep=0.75pt]  [font=\footnotesize]  {$\ket{\psi _{in}}$};
% Text Node
\draw (291.09,105.46) node [anchor=north west][inner sep=0.75pt]   [align=left] {{\fontfamily{ptm}\selectfont Region I}};

\end{tikzpicture}

    \caption{Penrose diagram of evaporating black hole.}
    \label{fig:my_label}
\end{figure}

Consider a mater distribution that collapses into a black hole and evoporates in a finite time. Suppose $\Sigma _-$ and $\Sigma_+$ are two Cauchy surfaces in the far past and the far future with respect to the black hole as shown in figure \ref{fig:my_label}. Cauchy surface $\Sigma_{BH}$ passes through the formal intersection of the horizon and the singularity in the Penrose diagram \ref{fig:my_label}. $\Sigma_{in}$ and $\Sigma_{out}$ denote the parts of $\Sigma_{BH}$ that are respectively inside and outside of the black hole. Suppose $\ket{\psi_-}$, $\ket{\psi_+}$, and $\ket{\psi_{BH}}$ are the state of the quantum fields respectively on $\Sigma_-$, $\Sigma_+$, and $\Sigma_{BH}$, and $\rho_{in}$ and $\rho_{out}$ are the density matrices associated to the quantum fields inside and outside the black hole given by 
\begin{align}
 &\rho_{in}=Tr_{\mathcal{H}_{out}}\ket{\psi_{BH}}\bra{\psi_{BH}},\nonumber\\
 &\rho_{out}=Tr_{\mathcal{H}_{in}}\ket{\psi_{BH}}\bra{\psi_{BH}}.\nonumber\\
\end{align}

From postulate 1, we know that some unitary transformation maps $\ket{\psi_-}$ to $\ket{\psi_+}$. 
\begin{equation}\label{BHC1}
\ket{\psi_+}=U_1\ket{\psi_-}.
\end{equation}
Similarly, since $\Sigma_+$ and $\Sigma_{out}$ are both Cauchy surfaces of region I in figure \ref{fig:my_label}, $\rho_{out}$ is related to $\ket{\psi_{+}}\bra{\psi_{+}}$ by a unitary transformation. Therefore, there must be a pure state $\ket{\psi_{out}}$ which satisfies $\rho_{out}=\ket{\psi_{out}}\bra{\psi_{out}}$ and is related to $\ket{\psi_+}$ via some unitary transformation $U_2$. 
\begin{equation}\label{BHC2}
\ket{\psi_{out}}=U_2\ket{\psi_+}.
\end{equation}

Combining \eqref{BHC1} and \eqref{BHC2} gives
\begin{align}
    \ket{\psi_{out}}=U_3\ket{\psi_-},
\end{align}
where $U_3=U_2U_1$. Postulate 1 tells us that $\ket{\psi_-}$ is unitarily mapped to $\ket{\psi_{BH}}$ as well. The only way $\ket{\psi_-}$ could be unitarily mapped to both $\ket{\psi_{BH}}$ and $\ket{\psi_{out}}$ is that $\ket{\psi_{BH}}=\ket{\psi_{in}}\otimes \ket{\psi_{out}}$ for some constant state $\ket{\psi_{in}}\in\mathcal{H}_{in}$. This however would mean that if a free falling observer falls into the black hole and hit $\Sigma_{in}$, they would see a fixed state $\ket{\psi_{in}}$ independent from the initial state $\ket{\psi_{-}}$. This is different from what an inertial observer would see in flat spacetime and therefore is a violation of the equivalence principle. 

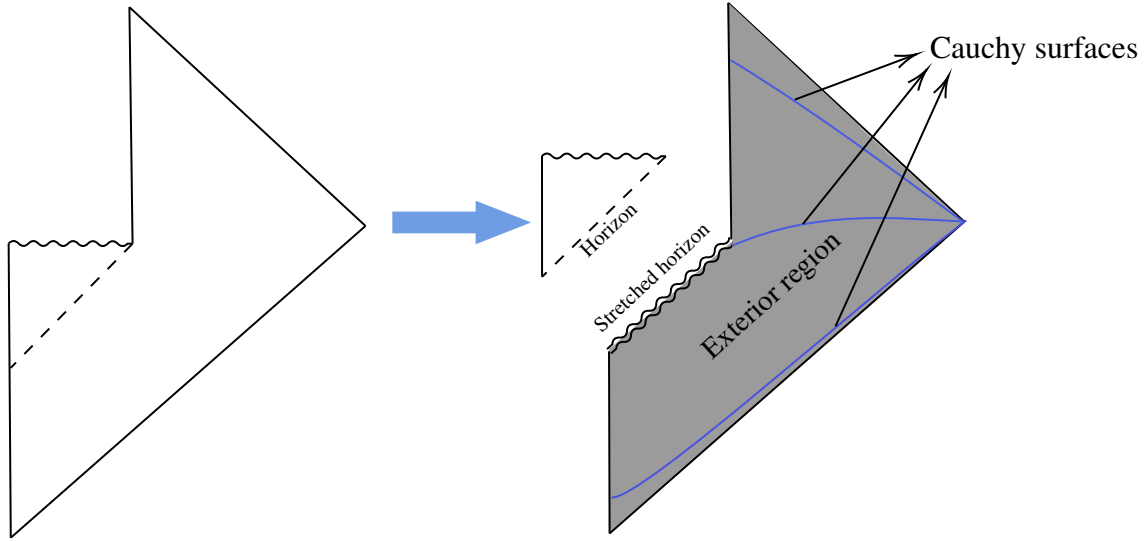
\begin{figure}
    \centering

\tikzset{every picture/.style={line width=0.75pt}} %set default line width to 0.75pt        

\begin{tikzpicture}[x=0.75pt,y=0.75pt,yscale=-1,xscale=1]
%uncomment if require: \path (0,329); %set diagram left start at 0, and has height of 329

%Straight Lines [id:da3058891685442342] 
\draw    (121.38,25.32) -- (240.46,135.76) ;
%Straight Lines [id:da4941925014319206] 
\draw    (60.47,144.59) -- (61.55,293.14) ;
%Straight Lines [id:da4576257552779064] 
\draw    (60.47,144.59) .. controls (62.14,142.93) and (63.81,142.93) .. (65.47,144.6) .. controls (67.14,146.27) and (68.8,146.27) .. (70.47,144.61) .. controls (72.14,142.95) and (73.8,142.95) .. (75.47,144.62) .. controls (77.14,146.29) and (78.8,146.29) .. (80.47,144.63) .. controls (82.14,142.97) and (83.8,142.97) .. (85.47,144.64) .. controls (87.14,146.31) and (88.8,146.31) .. (90.47,144.64) .. controls (92.14,142.98) and (93.8,142.98) .. (95.47,144.65) .. controls (97.14,146.32) and (98.8,146.32) .. (100.47,144.66) .. controls (102.14,143) and (103.8,143) .. (105.47,144.67) .. controls (107.14,146.34) and (108.8,146.34) .. (110.47,144.68) .. controls (112.14,143.02) and (113.8,143.02) .. (115.47,144.69) .. controls (117.14,146.36) and (118.8,146.36) .. (120.47,144.7) -- (122.98,144.7) -- (122.98,144.7) ;
%Straight Lines [id:da7907221027586944] 
\draw    (121.38,25.32) -- (122.98,144.7) ;
%Straight Lines [id:da8425065767458539] 
\draw    (240.46,135.76) -- (61.55,293.14) ;
%Straight Lines [id:da85897673293392] 
\draw  [dash pattern={on 4.5pt off 4.5pt}]  (122.98,144.7) -- (61.5,207.43) ;
%Right Arrow [id:dp7147431310996795] 
\draw  [color={rgb, 255:red, 126; green, 211; blue, 33 }  ,draw opacity=0 ][fill={rgb, 255:red, 2; green, 83; blue, 208 }  ,fill opacity=0.57 ] (254,128.5) -- (296,128.5) -- (296,123) -- (324,134) -- (296,145) -- (296,139.5) -- (254,139.5) -- cycle ;
%Shape: Trapezoid [id:dp35877184159794906] 
\draw  [color={rgb, 255:red, 0; green, 0; blue, 0 }  ,draw opacity=0 ][fill={rgb, 255:red, 155; green, 155; blue, 155 }  ,fill opacity=1 ] (363.55,291) -- (363.32,198.03) -- (451.27,120.76) -- (543.44,132.97) -- cycle ;
%Straight Lines [id:da6067011438455137] 
\draw    (423.38,23.18) -- (542.46,133.62) ;
%Shape: Triangle [id:dp41233702424181984] 
\draw  [color={rgb, 255:red, 0; green, 0; blue, 0 }  ,draw opacity=0 ][fill={rgb, 255:red, 155; green, 155; blue, 155 }  ,fill opacity=1 ] (490.47,85.85) -- (424.98,142.56) -- (423.38,23.18) -- cycle ;
%Straight Lines [id:da05860805412289927] 
\draw    (363.27,199.04) -- (363.55,291) ;
%Straight Lines [id:da889841022556829] 
\draw    (329.47,100.45) .. controls (331.14,98.78) and (332.81,98.79) .. (334.47,100.46) .. controls (336.14,102.13) and (337.8,102.13) .. (339.47,100.47) .. controls (341.14,98.8) and (342.8,98.8) .. (344.47,100.47) .. controls (346.14,102.14) and (347.8,102.14) .. (349.47,100.48) .. controls (351.14,98.82) and (352.8,98.82) .. (354.47,100.49) .. controls (356.14,102.16) and (357.8,102.16) .. (359.47,100.5) .. controls (361.14,98.84) and (362.8,98.84) .. (364.47,100.51) .. controls (366.14,102.18) and (367.8,102.18) .. (369.47,100.52) .. controls (371.14,98.86) and (372.8,98.86) .. (374.47,100.53) .. controls (376.14,102.2) and (377.8,102.2) .. (379.47,100.54) .. controls (381.14,98.88) and (382.8,98.88) .. (384.47,100.55) .. controls (386.14,102.22) and (387.8,102.22) .. (389.47,100.56) -- (391.98,100.56) -- (391.98,100.56) ;
%Straight Lines [id:da24463937544471315] 
\draw    (423.38,23.18) -- (424.98,142.56) ;
%Straight Lines [id:da13219165649765263] 
\draw    (542.46,133.62) -- (363.55,291) ;
%Straight Lines [id:da6269915997789508] 
\draw    (426,143.67) .. controls (425.89,146.02) and (424.66,147.15) .. (422.31,147.04) .. controls (419.96,146.94) and (418.73,148.07) .. (418.62,150.42) .. controls (418.51,152.77) and (417.28,153.9) .. (414.93,153.79) .. controls (412.58,153.69) and (411.35,154.82) .. (411.24,157.17) .. controls (411.13,159.52) and (409.9,160.65) .. (407.55,160.55) .. controls (405.2,160.44) and (403.97,161.57) .. (403.87,163.92) .. controls (403.76,166.27) and (402.53,167.4) .. (400.18,167.3) .. controls (397.83,167.19) and (396.6,168.32) .. (396.49,170.67) .. controls (396.38,173.02) and (395.15,174.15) .. (392.8,174.05) .. controls (390.45,173.94) and (389.22,175.07) .. (389.11,177.42) .. controls (389,179.77) and (387.77,180.9) .. (385.42,180.8) .. controls (383.07,180.7) and (381.84,181.83) .. (381.73,184.18) .. controls (381.63,186.53) and (380.4,187.66) .. (378.05,187.55) .. controls (375.7,187.45) and (374.47,188.58) .. (374.36,190.93) .. controls (374.25,193.28) and (373.02,194.41) .. (370.67,194.3) .. controls (368.32,194.2) and (367.09,195.33) .. (366.98,197.68) -- (364.28,200.15) -- (364.28,200.15)(423.97,141.45) .. controls (423.86,143.81) and (422.63,144.94) .. (420.28,144.83) .. controls (417.93,144.73) and (416.7,145.86) .. (416.59,148.21) .. controls (416.49,150.56) and (415.26,151.69) .. (412.91,151.58) .. controls (410.56,151.48) and (409.33,152.61) .. (409.22,154.96) .. controls (409.11,157.31) and (407.88,158.44) .. (405.53,158.33) .. controls (403.18,158.23) and (401.95,159.36) .. (401.84,161.71) .. controls (401.73,164.06) and (400.5,165.19) .. (398.15,165.08) .. controls (395.8,164.98) and (394.57,166.11) .. (394.46,168.46) .. controls (394.35,170.81) and (393.12,171.94) .. (390.77,171.84) .. controls (388.42,171.73) and (387.19,172.86) .. (387.09,175.21) .. controls (386.98,177.56) and (385.75,178.69) .. (383.4,178.59) .. controls (381.05,178.48) and (379.82,179.61) .. (379.71,181.96) .. controls (379.6,184.31) and (378.37,185.44) .. (376.02,185.34) .. controls (373.67,185.23) and (372.44,186.36) .. (372.33,188.71) .. controls (372.22,191.06) and (370.99,192.19) .. (368.64,192.09) .. controls (366.29,191.99) and (365.06,193.12) .. (364.96,195.47) -- (362.26,197.94) -- (362.26,197.94) ;
%Shape: Triangle [id:dp8907528395610227] 
\draw  [color={rgb, 255:red, 0; green, 0; blue, 0 }  ,draw opacity=0 ][fill={rgb, 255:red, 155; green, 155; blue, 155 }  ,fill opacity=1 ] (488.59,84.88) -- (541.83,133.53) -- (446.22,119.83) -- cycle ;
%Shape: Rectangle [id:dp12425212040687672] 
\draw  [color={rgb, 255:red, 155; green, 155; blue, 155 }  ,draw opacity=1 ][fill={rgb, 255:red, 155; green, 155; blue, 155 }  ,fill opacity=1 ] (426,117.86) -- (496,117.86) -- (496,157.86) -- (426,157.86) -- cycle ;
%Straight Lines [id:da20928267946919976] 
\draw    (329.47,100.45) -- (329.5,161.43) ;
%Straight Lines [id:da7025580451340718] 
\draw  [dash pattern={on 4.5pt off 4.5pt}]  (391.98,100.56) -- (329.5,161.43) ;
%Curve Lines [id:da08683002904075798] 
\draw [color={rgb, 255:red, 74; green, 81; blue, 226 }  ,draw opacity=1 ]   (364.48,272.78) .. controls (378.48,275.21) and (495.5,169.43) .. (543.44,132.97) ;
%Curve Lines [id:da5125993550164998] 
\draw [color={rgb, 255:red, 74; green, 91; blue, 226 }  ,draw opacity=1 ]   (424.5,52) .. controls (472.5,81) and (525.5,122) .. (541.83,133.53) ;
%Curve Lines [id:da5115425508286715] 
\draw [color={rgb, 255:red, 74; green, 91; blue, 226 }  ,draw opacity=1 ]   (425.75,145.6) .. controls (468.5,127) and (509.5,132) .. (541.83,133.53) ;
%Straight Lines [id:da05033567567715935] 
\draw    (457,72) -- (516.63,49.7) ;
\draw [shift={(518.5,49)}, rotate = 519.5] [color={rgb, 255:red, 0; green, 0; blue, 0 }  ][line width=0.75]    (10.93,-3.29) .. controls (6.95,-1.4) and (3.31,-0.3) .. (0,0) .. controls (3.31,0.3) and (6.95,1.4) .. (10.93,3.29)   ;
%Straight Lines [id:da5591141401069082] 
\draw    (461,134.86) -- (522.25,58.42) ;
\draw [shift={(523.5,56.86)}, rotate = 488.7] [color={rgb, 255:red, 0; green, 0; blue, 0 }  ][line width=0.75]    (10.93,-3.29) .. controls (6.95,-1.4) and (3.31,-0.3) .. (0,0) .. controls (3.31,0.3) and (6.95,1.4) .. (10.93,3.29)   ;
%Straight Lines [id:da48972993349302074] 
\draw    (477.5,186.43) -- (533.68,61.82) ;
\draw [shift={(534.5,60)}, rotate = 474.27] [color={rgb, 255:red, 0; green, 0; blue, 0 }  ][line width=0.75]    (10.93,-3.29) .. controls (6.95,-1.4) and (3.31,-0.3) .. (0,0) .. controls (3.31,0.3) and (6.95,1.4) .. (10.93,3.29)   ;

% Text Node
\draw (352.53,184.5) node [anchor=north west][inner sep=0.75pt]  [font=\scriptsize,rotate=-316.81] [align=left] {{\fontfamily{ptm}\selectfont Stretched horizon}};
% Text Node
\draw (345.59,146.57) node [anchor=north west][inner sep=0.75pt]  [font=\scriptsize,rotate=-316.51] [align=left] {{\fontfamily{ptm}\selectfont Horizon}};
% Text Node
\draw (523.7,38.82) node [anchor=north west][inner sep=0.75pt]   [align=left] {{\fontfamily{ptm}\selectfont Cauchy surfaces}};
% Text Node
\draw (406.49,195.96) node [anchor=north west][inner sep=0.75pt]  [rotate=-317.87] [align=left] {{\fontfamily{ptm}\selectfont {\small Exterior region}}};

\end{tikzpicture}

    \caption{The black hole complementarity allows us to isolate the black hole's exterior and study the evolution in it independently.}
    \label{fig:my_label2}
\end{figure}

The black hole complementarity is a principle to get around this paradox. The principle is that there are two different but complementary descriptions for the physics depending on the observer's trajectory. For outside observers at fixed distance from the black hole, the evolution could be studied independently from the interior of the black hole as shown in figure \ref{fig:my_label2}. All the black hole interactions with the exterior are explained by a real physical membrane at Planckian distance from the horizon. This membrane is called the \textit{stretched horizon}. The falling of matter inside the black hole could be viewed as the stretched horizon absorbing its energy. For outside observers, the Hawking radiation is the thermal radiation of the stretched horizon.

In contrast to the accelerating observers at a fixed distance from the black hole, a free-falling observer falling into the black hole will not see the stretched horizon. This leads to two different descriptions of the physical events on a given Cauchy surface that extends to the inside of the black hole. This might seem paradoxical at first; however, since the two observers are causally disconnected, they cannot communicate their different narratives to each other. In other words, no observer can experience both physics \footnote{Such an observer is often called a \textit{superobserver} in the literature. }. 

As we mentioned above, the black hole's Hawking radiation could be interpreted as the stretched horizon's thermal radiation. When an object falls in the black hole, it perturbs the stretched horizon and causes a small deviation from the equilibrium. After some time, the system's information thermalizes and can be radiated away in the form of thermal radiation. The time it takes for external perturbations to thermalize is called the \textit{scrambling time}. Following, we give a more rigorous definition of it and study it in more detail. 

The scrambling time of a system is the time that it takes for the information of a generic pure state to disperse among all microscopic degrees of freedom. More precisely, it is the time by which the density matrix $\rho=Tr_{\mathcal{H}_s}\ket\psi\bra\psi$ becomes thermal for almost every subsystem $\mathcal{H}_s$ with half the degrees of freedom. In \cite{Sekino:2008he}, it was conjectured that for any quantum system at inverse temperature $\beta$, the scrambling time $\tau_S$ is bounded from below by
\begin{align}\label{scrambling}
    \tau_s>f(\beta)\ln(S),
\end{align}
where $f(\beta)$ captures the interaction strength and $S$ is the total entropy. The systems that saturate the above bound for some function $f(\beta)$ are called \textit{fast scramblers}. Based on the complementarity principle, it was conjectured that both black holes and de Sitter space are fast scramblers with the scrambling time given by \cite{Sekino:2008he,Susskind:2011ap}
\begin{align}\label{FSE}
    \tau_s\sim \frac{1}{T}\ln(S).
\end{align}
It is worth taking a while to review the argument that bounds the black hole scrambling time from below. We review a thought experiment presented in \cite{Sekino:2008he} that shows if the scrambling time is too short, the no-cloning theorem could be violated. 

Suppose we have two observers Alice and Bob each carrying a q-bit that are fully entangled with each other. Bob jumps in the black hole, and right after he passes the horizon by a Planck length\footnote{smaller distances are not meaningful to a semi-classical observer.}, he measures the q-bit. Then, he sends a null signal in the outward radial direction that carries the measurement's outcome information. On the other side, Alice waits at a short distance outside the horizon until the black hole radiates away Bob's q-bit information. She then collects the Hawking radiation that contains Bob's q-bit's information\footnote{The experiment is done after the Page time so that Alice can retrieve Bob's q-bit's information with $\mathcal{O}(1)$ bits of Hawking radiation \cite{Page:1993wv,Hayden:2007cs}.}.

\begin{figure}[H]
    \centering

\tikzset{every picture/.style={line width=0.75pt}} %set default line width to 0.75pt        

\begin{tikzpicture}[x=0.75pt,y=0.75pt,yscale=-1,xscale=1]

%Straight Lines [id:da5767228664095327] 
\draw    (372.5,107.43) -- (371.5,11.43) ;
%Straight Lines [id:da9024766360344592] 
\draw [color={rgb, 255:red, 155; green, 155; blue, 155 }  ,draw opacity=1 ] [dash pattern={on 4.5pt off 4.5pt}]  (229.5,250.43) -- (372.5,107.43) ;
%Straight Lines [id:da5807823046320528] 
\draw    (230.5,292.43) -- (230,107) ;
%Curve Lines [id:da7709682079416491] 
\draw [color={rgb, 255:red, 74; green, 144; blue, 226 }  ,draw opacity=1 ]   (289.5,174.43) .. controls (293.5,180.43) and (296.5,181.43) .. (298.5,196.43) ;
%Straight Lines [id:da5608769769753088] 
\draw [color={rgb, 255:red, 74; green, 144; blue, 226 }  ,draw opacity=0.73 ]   (289.5,174.43) .. controls (289.43,172.07) and (290.58,170.86) .. (292.93,170.79) .. controls (295.28,170.72) and (296.42,169.5) .. (296.35,167.15) .. controls (296.28,164.8) and (297.43,163.58) .. (299.78,163.51) .. controls (302.14,163.44) and (303.28,162.22) .. (303.21,159.86) .. controls (303.14,157.51) and (304.28,156.29) .. (306.63,156.22) .. controls (308.98,156.15) and (310.13,154.93) .. (310.06,152.58) .. controls (309.99,150.23) and (311.14,149.01) .. (313.49,148.94) .. controls (315.84,148.87) and (316.98,147.65) .. (316.91,145.3) .. controls (316.84,142.95) and (317.99,141.73) .. (320.34,141.66) .. controls (322.69,141.59) and (323.84,140.37) .. (323.77,138.02) .. controls (323.7,135.67) and (324.85,134.45) .. (327.2,134.38) .. controls (329.55,134.31) and (330.69,133.09) .. (330.62,130.74) .. controls (330.55,128.39) and (331.7,127.17) .. (334.05,127.1) .. controls (336.41,127.03) and (337.55,125.81) .. (337.48,123.45) .. controls (337.41,121.1) and (338.55,119.88) .. (340.9,119.81) .. controls (343.25,119.74) and (344.4,118.52) .. (344.33,116.17) .. controls (344.26,113.82) and (345.41,112.6) .. (347.76,112.53) .. controls (350.11,112.46) and (351.25,111.24) .. (351.18,108.89) -- (353.5,106.43) -- (353.5,106.43) ;
%Curve Lines [id:da08427603802510508] 
\draw [color={rgb, 255:red, 208; green, 2; blue, 27 }  ,draw opacity=1 ]   (298.5,196.43) .. controls (313.5,175.43) and (339.5,147.43) .. (366.5,119.43) ;
%Straight Lines [id:da06480613226825693] 
\draw [color={rgb, 255:red, 208; green, 2; blue, 27 }  ,draw opacity=1 ]   (353.5,106.43) -- (366.5,119.43) ;
%Shape: Circle [id:dp9887430690200831] 
\draw  [draw opacity=0][fill={rgb, 255:red, 255; green, 255; blue, 255 }  ,fill opacity=1 ] (349.21,106.43) .. controls (349.21,104.06) and (351.13,102.14) .. (353.5,102.14) .. controls (355.87,102.14) and (357.79,104.06) .. (357.79,106.43) .. controls (357.79,108.8) and (355.87,110.71) .. (353.5,110.71) .. controls (351.13,110.71) and (349.21,108.8) .. (349.21,106.43) -- cycle ;
%Straight Lines [id:da816160591839926] 
\draw    (230,107) .. controls (231.67,105.34) and (233.34,105.35) .. (235,107.02) .. controls (236.67,108.69) and (238.33,108.69) .. (240,107.03) .. controls (241.67,105.37) and (243.34,105.38) .. (245,107.05) .. controls (246.67,108.72) and (248.33,108.72) .. (250,107.06) .. controls (251.67,105.4) and (253.34,105.41) .. (255,107.08) .. controls (256.67,108.75) and (258.33,108.75) .. (260,107.09) .. controls (261.67,105.43) and (263.34,105.44) .. (265,107.11) .. controls (266.67,108.78) and (268.33,108.78) .. (270,107.12) .. controls (271.67,105.46) and (273.34,105.47) .. (275,107.14) .. controls (276.67,108.81) and (278.33,108.81) .. (280,107.15) .. controls (281.67,105.49) and (283.34,105.5) .. (285,107.17) .. controls (286.67,108.84) and (288.33,108.84) .. (290,107.18) .. controls (291.67,105.52) and (293.34,105.53) .. (295,107.2) .. controls (296.67,108.87) and (298.33,108.87) .. (300,107.21) .. controls (301.67,105.55) and (303.34,105.56) .. (305,107.23) .. controls (306.67,108.9) and (308.33,108.9) .. (310,107.24) .. controls (311.67,105.58) and (313.34,105.59) .. (315,107.26) .. controls (316.67,108.93) and (318.33,108.93) .. (320,107.27) .. controls (321.67,105.61) and (323.34,105.62) .. (325,107.29) .. controls (326.67,108.96) and (328.33,108.96) .. (330,107.3) .. controls (331.67,105.64) and (333.34,105.65) .. (335,107.32) .. controls (336.67,108.99) and (338.33,108.99) .. (340,107.33) .. controls (341.67,105.67) and (343.34,105.68) .. (345,107.35) .. controls (346.67,109.02) and (348.33,109.02) .. (350,107.36) .. controls (351.67,105.7) and (353.34,105.71) .. (355,107.38) .. controls (356.67,109.05) and (358.33,109.05) .. (360,107.39) .. controls (361.67,105.73) and (363.34,105.74) .. (365,107.41) .. controls (366.67,109.08) and (368.33,109.08) .. (370,107.42) -- (372.5,107.43) -- (372.5,107.43) ;

% Text Node
\draw (262,175) node [anchor=north west][inner sep=0.75pt]  [font=\small] [align=left] {{\fontfamily{ptm}\selectfont {\small Bob}}};
% Text Node
\draw (331,158) node [anchor=north west][inner sep=0.75pt]   [align=left] {{\fontfamily{ptm}\selectfont {\small Alice}}};
% Text Node
\draw (278.47,161.32) node [anchor=north west][inner sep=0.75pt]  [font=\small,rotate=-312.12] [align=left] {{\fontfamily{ptm}\selectfont {\footnotesize Bob's signal}}};

\end{tikzpicture}

    \caption{If Alice's worldline (the red curve) could meet Bob's signal (blue wavy curve), no-cloning theorem would be violated.}
    \label{nctv}
\end{figure}
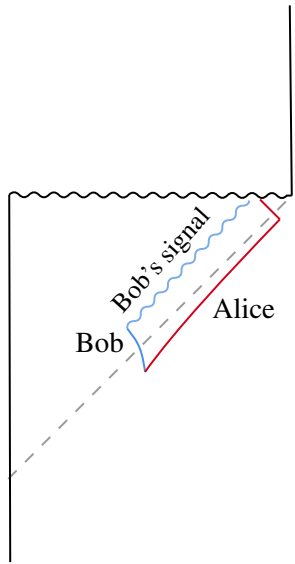

The time Alice needs to wait is precisely the scrambling time. As soon as Alice receives a copy of Bob's q-bit's information through Hawking radiation, she jumps in to catch Bob's signal (figure \ref{nctv}). If she succeeds, she will have two copies of Bob's information, which violates the no-cloning theorem. Moreover, both copies are fully entangled with Alice's q-bit, which violates the monogamy theorem. The scrambling time must be long enough so that Alice's future light cone and Bob's signal do not intersect to avoid these contradictions. This gives a lower bound of $\sim \frac{1}{M}\log(M)$ for the scrambling time \cite{Sekino:2008he}.

In the next subsection, we apply the ideas we covered in this subsection to de Sitter space to develop the de Sitter complementarity. 

\subsection{de Sitter complementarity}\label{DSC1}

\begin{figure}
    \centering

\tikzset{every picture/.style={line width=0.75pt}} %set default line width to 0.75pt        

\begin{tikzpicture}[x=0.75pt,y=0.75pt,yscale=-1,xscale=1]
%uncomment if require: \path (0,339); %set diagram left start at 0, and has height of 339

%Shape: Triangle [id:dp2783729510991213] 
\draw  [color={rgb, 255:red, 0; green, 0; blue, 0 }  ,draw opacity=0 ][fill={rgb, 255:red, 236; green, 206; blue, 206 }  ,fill opacity=1 ] (197.76,37.24) -- (198,19) -- (286.64,118.67) -- cycle ;
%Shape: Rectangle [id:dp7326740526276638] 
\draw   (198,19) -- (488.5,19) -- (488.5,299.33) -- (198,299.33) -- cycle ;
%Straight Lines [id:da40380951784729757] 
\draw    (198,19) -- (488.5,299.33) ;
%Curve Lines [id:da3126379933368546] 
\draw [color={rgb, 255:red, 46; green, 79; blue, 226 }  ,draw opacity=1 ] [dash pattern={on 4.5pt off 4.5pt}]  (198,19) .. controls (250.5,70.57) and (304.59,128.54) .. (297.59,149.54) ;
%Curve Lines [id:da5508960786919315] 
\draw    (297.59,149.54) .. controls (299.39,129.5) and (305.75,117.68) .. (315.72,104.35) ;
\draw [shift={(317.5,102)}, rotate = 487.5] [fill={rgb, 255:red, 0; green, 0; blue, 0 }  ][line width=0.08]  [draw opacity=0] (10.72,-5.15) -- (0,0) -- (10.72,5.15) -- (7.12,0) -- cycle    ;
%Straight Lines [id:da6998096629322441] 
\draw    (306.58,153.6) -- (319.5,141.33) ;
\draw [shift={(319.5,141.33)}, rotate = 496.49] [color={rgb, 255:red, 0; green, 0; blue, 0 }  ][line width=0.75]    (0,5.59) -- (0,-5.59)   ;
\draw [shift={(306.58,153.6)}, rotate = 496.49] [color={rgb, 255:red, 0; green, 0; blue, 0 }  ][line width=0.75]    (0,5.59) -- (0,-5.59)   ;
%Shape: Star [id:dp601948801606724] 
\draw  [color={rgb, 255:red, 0; green, 0; blue, 0 }  ,draw opacity=0 ][fill={rgb, 255:red, 236; green, 107; blue, 107 }  ,fill opacity=1 ] (288.04,113.83) -- (289.59,117.57) -- (293.04,118.17) -- (290.54,121.09) -- (291.13,125.2) -- (288.04,123.26) -- (284.96,125.2) -- (285.55,121.09) -- (283.05,118.17) -- (286.5,117.57) -- cycle ;
%Shape: Star [id:dp4635738187448777] 
\draw  [color={rgb, 255:red, 0; green, 0; blue, 0 }  ,draw opacity=0 ][fill={rgb, 255:red, 107; green, 135; blue, 236 }  ,fill opacity=1 ] (297.59,143.26) -- (299.13,147) -- (302.58,147.6) -- (300.08,150.51) -- (300.67,154.63) -- (297.59,152.68) -- (294.5,154.63) -- (295.09,150.51) -- (292.59,147.6) -- (296.04,147) -- cycle ;
%Straight Lines [id:da17960694008543232] 
\draw    (285.5,150.33) -- (279.04,129.11) ;
\draw [shift={(279.04,129.11)}, rotate = 433.07] [color={rgb, 255:red, 0; green, 0; blue, 0 }  ][line width=0.75]    (0,5.59) -- (0,-5.59)   ;
\draw [shift={(285.5,150.33)}, rotate = 433.07] [color={rgb, 255:red, 0; green, 0; blue, 0 }  ][line width=0.75]    (0,5.59) -- (0,-5.59)   ;
%Straight Lines [id:da10038536610715365] 
\draw    (198,299.33) -- (488.5,19) ;
%Shape: Triangle [id:dp4847353379765251] 
\draw  [color={rgb, 255:red, 0; green, 0; blue, 0 }  ,draw opacity=0 ][fill={rgb, 255:red, 236; green, 206; blue, 206 }  ,fill opacity=1 ] (247.32,82.58) -- (250.21,73.86) -- (281.18,112.55) -- cycle ;
%Flowchart: Alternative Process [id:dp34983383099245313] 
\draw  [draw opacity=0][fill={rgb, 255:red, 236; green, 206; blue, 206 }  ,fill opacity=1 ] (212.43,35.57) .. controls (212.86,35.18) and (213.54,35.21) .. (213.93,35.64) -- (263.89,90.38) .. controls (264.29,90.81) and (264.26,91.48) .. (263.82,91.88) -- (260.91,94.54) .. controls (260.47,94.94) and (259.8,94.91) .. (259.41,94.47) -- (209.44,39.74) .. controls (209.05,39.3) and (209.08,38.63) .. (209.51,38.24) -- cycle ;
%Straight Lines [id:da5688490219410769] 
\draw [color={rgb, 255:red, 226; green, 95; blue, 93 }  ,draw opacity=1 ]   (199.98,39.25) -- (205.91,44.62) .. controls (208.27,44.51) and (209.51,45.63) .. (209.62,47.98) .. controls (209.74,50.33) and (210.98,51.45) .. (213.33,51.33) .. controls (215.68,51.22) and (216.92,52.34) .. (217.03,54.69) .. controls (217.15,57.04) and (218.39,58.16) .. (220.74,58.04) .. controls (223.09,57.93) and (224.33,59.05) .. (224.45,61.4) .. controls (224.56,63.75) and (225.8,64.87) .. (228.15,64.75) .. controls (230.5,64.64) and (231.74,65.76) .. (231.86,68.11) .. controls (231.98,70.46) and (233.22,71.58) .. (235.57,71.46) .. controls (237.92,71.35) and (239.16,72.47) .. (239.27,74.82) .. controls (239.39,77.17) and (240.63,78.29) .. (242.98,78.18) .. controls (245.33,78.06) and (246.57,79.18) .. (246.69,81.53) .. controls (246.8,83.88) and (248.04,85) .. (250.39,84.89) .. controls (252.74,84.77) and (253.98,85.89) .. (254.1,88.24) .. controls (254.22,90.59) and (255.46,91.71) .. (257.81,91.6) .. controls (260.16,91.48) and (261.4,92.6) .. (261.51,94.95) .. controls (261.63,97.3) and (262.87,98.42) .. (265.22,98.31) .. controls (267.57,98.19) and (268.81,99.31) .. (268.93,101.66) .. controls (269.05,104.01) and (270.29,105.13) .. (272.64,105.02) .. controls (274.99,104.91) and (276.23,106.03) .. (276.34,108.38) .. controls (276.46,110.73) and (277.7,111.85) .. (280.05,111.73) .. controls (282.4,111.62) and (283.64,112.74) .. (283.76,115.09) -- (286.5,117.57) -- (286.5,117.57) ;
\draw [shift={(197.76,37.24)}, rotate = 42.15] [fill={rgb, 255:red, 226; green, 95; blue, 93 }  ,fill opacity=1 ][line width=0.08]  [draw opacity=0] (10.72,-5.15) -- (0,0) -- (10.72,5.15) -- (7.12,0) -- cycle    ;

% Text Node
\draw (233.09,21.98) node [anchor=north west][inner sep=0.75pt]  [rotate=-43.82] [align=left] {{\fontfamily{ptm}\selectfont Hubble Horizon}};
% Text Node
\draw (303.86,165.99) node [anchor=north west][inner sep=0.75pt]  [font=\scriptsize,rotate=-317.14]  {$\sim l_{pl}$};
% Text Node
\draw (316,84) node [anchor=north west][inner sep=0.75pt]   [align=left] {{\fontfamily{ptm}\selectfont B}};
% Text Node
\draw (180,34) node [anchor=north west][inner sep=0.75pt]   [align=left] {{\fontfamily{ptm}\selectfont A}};
% Text Node
\draw (238.9,134.82) node [anchor=north west][inner sep=0.75pt]  [font=\scriptsize,rotate=-358.81]  {${\textstyle \Delta t\sim \tau _{s}}$};

\end{tikzpicture}

    \caption{Penrose diagram of de Sitter space. The left side is the world line of a comoving observer A and the dashed blue curve denotes its corresponding stretched horizon. B is a system that exits A's Hubble horizon. The information of B thermalizes over the stretched horizon after the scrambling time $\tau_s$ and is partially radiated by the Hawking radiation in the red region.}
    \label{fig:my_label3}
\end{figure}
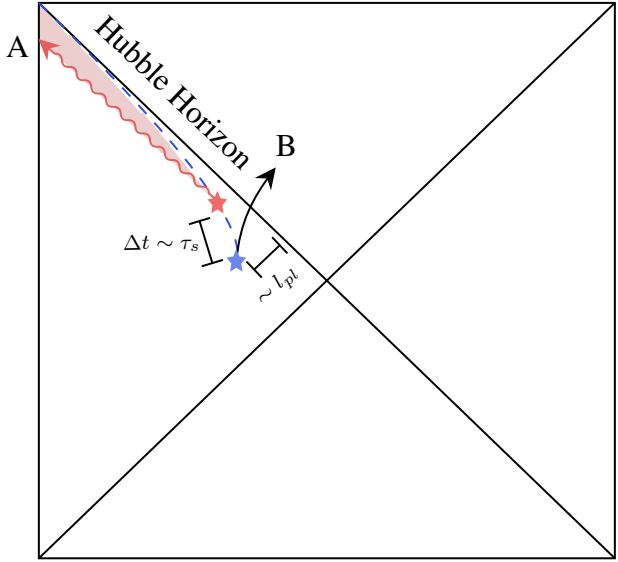

Several authors \cite{tw1,banksb,fisch,tw2,boussoN, lisa,birthday,lisamatt,eric} have proposed a complementarity principle for de Sitter space similar to the black hole version that we discussed in the previous subsection. The proposal is that the physics inside the Hubble horizon could be described independently from the outside of the Hubble patch. Similar to the black hole version, there is a stretched horizon located a Planck distance from the Hubble horizon. The stretched horizon disappears from free-falling observers crossing the Hubble horizon. When a system crosses the Hubble horizon of a comoving observer, the information of that system thermalizes over the stretched horizon and gets radiated toward the observer after a scrambling time (see figure \ref{fig:my_label3}). 

A key difference between black hole complementarity and its de Sitter counterpart is in information recovery. In order to recover the information of a system that has crossed the black hole's horizon, one needs to wait out the Hawking-Page transition \cite{Page:1993wv}. This is to ensure that half of the black hole's entropy is radiated away so that Hayden-Perskill protocol \cite{Hayden:2007cs} could be implemented. Another way to think about the Page time is when the radiated matter is maximally entangled with the remaining of the black hole. The significant role of the maximal entanglement is especially evident in ER=EPR duality \cite{Maldacena:2013xja}. Suppose one collapses the maximally entangled radiated matter into a second black hole. It was conjectured that this would create a wormhole geometry, which makes the information recovery possible \cite{Gao:2016bin,Maldacena:2017axo}. 

Information recovery is a bit trickier in de Sitter space. For starters, there is no Hawking-Page transition in de Sitter space. This is due to the fact that the maximum entropy that can be stored in the Hubble patch is a third of de Sitter entropy \cite{Parikh:2008iu}. At first, this might seem to suggest that information in de Sitter space is irretrievable. However, collecting half the entropy is unnecessary as long as we have access to a maximally entangled state with the stretched horizon. In that case, the information can be recovered as soon as it is scrambled. As we will discuss in more detail in subsection \ref{3p3}, after the scrambling time, the vacuum evolves into the Bunch-Davies vacuum, which is maximally entangled across the horizon. Therefore, after the scrambling time, all the necessary ingredients to recover information are in place. Authors in \cite{Aalsma:2020aib} present an elegant method to recover information after scrambling time with the use of shockwaves\footnote{By computing out-of-time-order correlators, they show that de Sitter space is a fast scrambler ($\tau_s\sim\frac{1}{H}\ln(\frac{1}{H})$). A similar argument for fast scrambling in de Sitter space was discussed in \cite{Geng:2020kxh}.}. 

For de Sitter space, the conjectured value \eqref{FSE} for the scrambling time takes the form 
\begin{align}\label{dsscr}
    \tau_s\sim\frac{1}{H}\ln(\frac{1}{H}),
\end{align}
where $H\propto \sqrt{\Lambda}$ is the Hubble parameter. As pointed out earlier, the scrambling time matches TCC time $\tau_{TCC}$. In this section we focus on de Sitter scrambling time and try to understand \eqref{dsscr} better. We try to find an argument similar to the thought experiment we mentioned for black holes that justifies \eqref{dsscr}. In Appendix \ref{31}, we present a thought experiment analogous to the one we discussed for black holes. We show that $\tau_s$ must be greater than $\sim\frac{1}{H}\ln(\frac{1}{H})$ to avoid a violation of the no-cloning principle. Here, we propose a different thought experiment that offers a clear insight into the relation between the scrambling time and the maximum lifetime from TCC.
\vspace{10pt}

\textbf{Thought experiment: Observational consistency}

In the framework of complementarity, different observers experience different physics. The comoving observers see a stretched horizon while the free-falling observers crossing the Hubble horizon do not. This is consistent as long as observers who experience different physics cannot communicate their different narratives to each other. This is trivially satisfied for black holes, but in de Sitter space, the situation is more tricky. We consider a thought experiment to check if observers who experience different physics can communicate their experience to each other. We show that the consistency of complementarity imposes a lower bound $\frac{1}{H}\ln(\frac{1}{H})$ on the scrambling time. 

\begin{figure}[H]
    \centering

\tikzset{every picture/.style={line width=0.75pt}} %set default line width to 0.75pt        

\begin{tikzpicture}[x=0.75pt,y=0.75pt,yscale=-1,xscale=1]
%uncomment if require: \path (0,223); %set diagram left start at 0, and has height of 223

%Shape: Star [id:dp44241157122965546] 
\draw  [color={rgb, 255:red, 0; green, 0; blue, 0 }  ,draw opacity=0 ][fill={rgb, 255:red, 107; green, 236; blue, 121 }  ,fill opacity=1 ] (265,21.03) -- (266.55,24.77) -- (270,25.37) -- (267.5,28.29) -- (268.09,32.4) -- (265,30.46) -- (261.92,32.4) -- (262.51,28.29) -- (260.01,25.37) -- (263.46,24.77) -- cycle ;
%Shape: Rectangle [id:dp5739086975036822] 
\draw  [color={rgb, 255:red, 0; green, 0; blue, 0 }  ,draw opacity=0 ] (190,-56) -- (478.5,-56) -- (478.5,196.57) -- (190,196.57) -- cycle ;
%Straight Lines [id:da1676577705772233] 
\draw    (190,-56) -- (458.5,181.29) ;
%Straight Lines [id:da26331925178339644] 
\draw [color={rgb, 255:red, 226; green, 95; blue, 93 }  ,draw opacity=1 ]   (269.76,30.26) -- (275.79,35.52) .. controls (278.14,35.36) and (279.39,36.46) .. (279.55,38.81) .. controls (279.71,41.16) and (280.97,42.26) .. (283.32,42.1) .. controls (285.67,41.94) and (286.93,43.04) .. (287.08,45.39) .. controls (287.24,47.74) and (288.5,48.84) .. (290.85,48.68) .. controls (293.2,48.51) and (294.46,49.61) .. (294.62,51.96) .. controls (294.77,54.31) and (296.03,55.41) .. (298.38,55.25) .. controls (300.73,55.09) and (301.99,56.19) .. (302.15,58.54) .. controls (302.31,60.89) and (303.57,61.99) .. (305.92,61.83) .. controls (308.27,61.67) and (309.53,62.77) .. (309.68,65.12) .. controls (309.84,67.47) and (311.1,68.57) .. (313.45,68.41) .. controls (315.8,68.25) and (317.06,69.35) .. (317.21,71.7) .. controls (317.37,74.05) and (318.63,75.15) .. (320.98,74.99) .. controls (323.33,74.82) and (324.59,75.92) .. (324.75,78.27) .. controls (324.9,80.62) and (326.16,81.72) .. (328.51,81.56) .. controls (330.86,81.4) and (332.12,82.5) .. (332.28,84.85) .. controls (332.43,87.2) and (333.69,88.3) .. (336.04,88.14) .. controls (338.39,87.98) and (339.65,89.08) .. (339.81,91.43) .. controls (339.97,93.78) and (341.23,94.88) .. (343.58,94.72) .. controls (345.93,94.56) and (347.19,95.66) .. (347.34,98.01) .. controls (347.5,100.36) and (348.76,101.46) .. (351.11,101.3) .. controls (353.46,101.13) and (354.72,102.23) .. (354.88,104.58) .. controls (355.03,106.93) and (356.29,108.03) .. (358.64,107.87) .. controls (360.99,107.71) and (362.25,108.81) .. (362.41,111.16) -- (363.53,112.14) -- (363.53,112.14) ;
\draw [shift={(267.5,28.29)}, rotate = 41.13] [fill={rgb, 255:red, 226; green, 95; blue, 93 }  ,fill opacity=1 ][line width=0.08]  [draw opacity=0] (10.72,-5.15) -- (0,0) -- (10.72,5.15) -- (7.12,0) -- cycle    ;
%Curve Lines [id:da9593076240678806] 
\draw [color={rgb, 255:red, 46; green, 79; blue, 226 }  ,draw opacity=1 ] [dash pattern={on 4.5pt off 4.5pt}]  (190,-56) .. controls (242.5,-4.43) and (413.49,148.51) .. (406.49,169.51) ;
%Curve Lines [id:da808715400275897] 
\draw    (406.5,187.57) .. controls (415.41,128.88) and (355.71,73.41) .. (287.57,16.03) ;
\draw [shift={(285.5,14.29)}, rotate = 400.05] [fill={rgb, 255:red, 0; green, 0; blue, 0 }  ][line width=0.08]  [draw opacity=0] (10.72,-5.15) -- (0,0) -- (10.72,5.15) -- (7.12,0) -- cycle    ;
%Straight Lines [id:da1360063665844855] 
\draw    (416.85,174.71) -- (431.15,163.43) ;
\draw [shift={(433.5,161.57)}, rotate = 501.71] [fill={rgb, 255:red, 0; green, 0; blue, 0 }  ][line width=0.08]  [draw opacity=0] (8.93,-4.29) -- (0,0) -- (8.93,4.29) -- cycle    ;
\draw [shift={(414.5,176.57)}, rotate = 321.71] [fill={rgb, 255:red, 0; green, 0; blue, 0 }  ][line width=0.08]  [draw opacity=0] (8.93,-4.29) -- (0,0) -- (8.93,4.29) -- cycle    ;
%Shape: Star [id:dp8534774872920463] 
\draw  [color={rgb, 255:red, 0; green, 0; blue, 0 }  ,draw opacity=0 ][fill={rgb, 255:red, 236; green, 107; blue, 107 }  ,fill opacity=1 ] (364.04,105.89) -- (365.58,109.63) -- (369.03,110.23) -- (366.53,113.14) -- (367.12,117.26) -- (364.04,115.31) -- (360.95,117.26) -- (361.54,113.14) -- (359.04,110.23) -- (362.49,109.63) -- cycle ;
%Shape: Star [id:dp9228642237778111] 
\draw  [color={rgb, 255:red, 0; green, 0; blue, 0 }  ,draw opacity=0 ][fill={rgb, 255:red, 107; green, 135; blue, 236 }  ,fill opacity=1 ] (406.49,163.23) -- (408.04,166.97) -- (411.49,167.57) -- (408.99,170.49) -- (409.58,174.6) -- (406.49,172.66) -- (403.41,174.6) -- (404,170.49) -- (401.5,167.57) -- (404.95,166.97) -- cycle ;
%Straight Lines [id:da8043906669222778] 
\draw    (398.32,170.45) -- (354.59,125.44) ;
\draw [shift={(352.5,123.29)}, rotate = 405.83000000000004] [fill={rgb, 255:red, 0; green, 0; blue, 0 }  ][line width=0.08]  [draw opacity=0] (8.93,-4.29) -- (0,0) -- (8.93,4.29) -- cycle    ;
\draw [shift={(400.41,172.6)}, rotate = 225.83] [fill={rgb, 255:red, 0; green, 0; blue, 0 }  ][line width=0.08]  [draw opacity=0] (8.93,-4.29) -- (0,0) -- (8.93,4.29) -- cycle    ;
%Straight Lines [id:da13931242809529776] 
\draw    (299.5,10.29) -- (481.5,170.29) ;
%Curve Lines [id:da4669793025742761] 
\draw  [dash pattern={on 4.5pt off 4.5pt}]  (191.5,63.29) .. controls (266.5,66.29) and (320,89.37) .. (364.04,112.17) ;
%Curve Lines [id:da35652903856159424] 
\draw  [dash pattern={on 4.5pt off 4.5pt}]  (191.5,154.29) .. controls (223.5,151.29) and (379.5,155.29) .. (406.49,169.51) ;
%Straight Lines [id:da9321866007152313] 
\draw    (221.5,66.29) -- (221.5,154.29) ;
%Straight Lines [id:da5316660540543727] 
\draw    (265,27.31) -- (221.5,66.29) ;
%Shape: Rectangle [id:dp6480619092818636] 
\draw  [color={rgb, 255:red, 0; green, 0; blue, 0 }  ,draw opacity=0 ][fill={rgb, 255:red, 255; green, 255; blue, 255 }  ,fill opacity=1 ] (160.5,-57.71) -- (283,-57.71) -- (283,23) -- (160.5,23) -- cycle ;
%Straight Lines [id:da9886411429212882] 
\draw    (191.5,63.29) -- (191.5,154.29) ;

% Text Node
\draw (416.1,184.18) node [anchor=north west][inner sep=0.75pt]  [font=\scriptsize,rotate=-324.49]  {${\textstyle \sim l_{P}}$};
% Text Node
\draw (206.14,122.09) node [anchor=north west][inner sep=0.75pt]  [rotate=-269.39] [align=left] {{\fontfamily{ptm}\selectfont Bob}};
% Text Node
\draw (174.74,126.62) node [anchor=north west][inner sep=0.75pt]  [rotate=-269.21] [align=left] {{\fontfamily{ptm}\selectfont Alice}};
% Text Node
\draw (334.9,143.82) node [anchor=north west][inner sep=0.75pt]  [font=\scriptsize,rotate=-358.81]  {${\textstyle \Delta t\sim \tau _{s}}$};
% Text Node
\draw (393,182) node [anchor=north west][inner sep=0.75pt]   [align=left] {{\fontfamily{ptm}\selectfont C}};
% Text Node
\draw (393.45,72.92) node [anchor=north west][inner sep=0.75pt]  [rotate=-40.94] [align=left] {{\fontfamily{ptm}\selectfont {\footnotesize Bob's Horizon}}};
% Text Node
\draw (418.45,124.92) node [anchor=north west][inner sep=0.75pt]  [rotate=-40.94] [align=left] {{\fontfamily{ptm}\selectfont {\footnotesize Alice's Horizon}}};

\end{tikzpicture}

    \caption{Setup of the thought experiment. System C exits Alice's Hubble horizon but stays inside Bob's Hubble patch. The dashed blue line is Alice's stretched horizon and the red squiggly line is the Hawking radiation which carries part of C's information.}
    \label{fig:my_label4}
\end{figure}
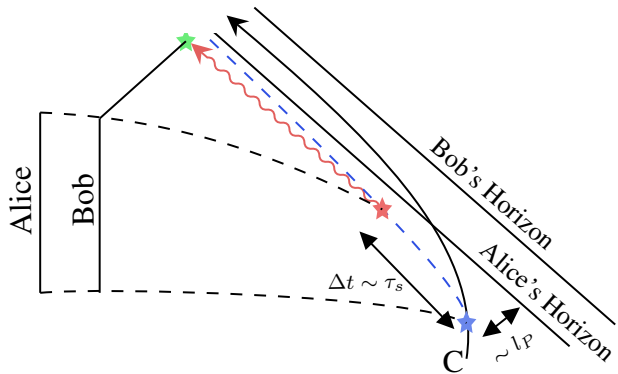

Consider two comoving observers Alice and Bob with initial positions of $r_A=0$ and $r_B=l$ at $t=0$. Consider a q-bit crosses Alice's horizon at $t=0$, but it accelerates afterwards such that it never crosses Bob's horizon. Suppose Alice and Bob continue to stay on their comoving paths until $t=\tau_s$. From Alice's perspective, Hawking radiation carrying part of (even if very small) the q-bit's information radiates inward. However, from Bob's point of view, that would not happen since the q-bit has never exited his horizon. If Bob is still within Alice's Hubble horizon at this time, Alice can catch the radiation with the q-bit's information and communicate her different narrative to Bob (see figure \ref{fig:my_label4}). Alice and Bob must have exited each other's Hubble horizon by the scrambling time to prevent these inconsistent narratives from ever meeting each other. Therefore, scrambling time should be longer than the time it takes for a comoving length $l$ to stretch beyond the Hubble radius.

\begin{align}
    \tau_s>\frac{1}{H}\ln(\frac{1}{Hl}).
\end{align}

We set $l$ to the smallest possible meaningful distance $l_{min}$. 
\begin{align}
    \tau_s>\frac{1}{H}\ln(\frac{1}{Hl_{min}}).
\end{align}
If $l_{min}=l_{P}$, the above lower bound matches TCC time. This thought experiment elucidates the relation between scrambling time and TCC time. They are both the time it takes for Planckian lengths to stretch beyond the Hubble radius. What is nice about this argument is that it tells us that the two times would match even if the smallest physical length scale $l_{min}$ were different from Planck length\footnote{For example, the size of the compact dimensions could set a greater lower bound for physical lengths.}. It is reasonable to believe that in such a background, TCC time scale as the maximum lifetime of de Sitter space gets replaced with $\frac{1}{H}\ln(\frac{1}{Hl_{min}})$. This is because the fundamental idea behind TCC time scale is to ensure that the smallest physical quantum fluctuations do not exit the Hubble horizon and classicalize\footnote{We thank Matthew Reece for sharing this insight.}. 

\subsection{Thermalization in de Sitter space}\label{3p3}

In the previous subsection, we studied thermalization in the framework of complementarity and showed that it takes $t\sim\frac{1}{H}\ln(\frac{1}{H})$ to happen. In this subsection, we support that result by studying the de Sitter vacua. We review quantum backgrounds with approximate de Sitter symmetries (de Sitter vacua) and an argument that shows in a time of order $\tau_{TCC}$ all of them evolve into a particular one called the \textit{Bunch-Davies} (BD) vacuum. The BD vacuum is a thermal background, which is why we call this process thermalization. 

If we have a de Sitter vacuum $\ket\Omega$ the Wightman two-point functions $\bra\Omega\phi(x)\phi(y)\ket\Omega$ must respect the isometries of the de Sitter space. Additionally, they are Green's functions of the free scalar field equation of motion. All such functions can be parametrized by a complex number $\alpha$ with a negative real part\footnote{See \cite{Allen:1985ux} for the analytic expression of the two-point function $G_\alpha$.}. The states $\ket\alpha$ for which 
\begin{align}
    G_\alpha(x,y)=\bra\alpha\phi(x)\phi(y)\ket\alpha,
\end{align}
are called $\alpha$-vacua. For $\alpha=-\infty$, the Green's function matches the thermal Green's function \cite{Allen:1985ux,Birrell:1982ix}. This vacuum is called the Bunch-Davis vacuum. A review of the mode expansions that lead to these vacua can be found in \cite{Mottola:1984ar}. The modes $\phi^{\pm}(k)$ corresponding to the BD vacuum have a special property that
\begin{align}\label{LMM}
    as~~k\rightarrow\infty~~\Rightarrow~~\phi^\pm(k)\sim\frac{H\eta^\frac{1}{2}}{k}e^{i\vec k\cdot\vec x\mp ik\eta},
\end{align}
where $(\vec x,\eta)$ are comoving coordinates. Equation \eqref{LMM} tells us that at large momenta (short distances) the BD modes exhibit the same behavior as their Minkowski counterparts. Therefore, the BD vacuum looks like the Minkowski vacuum at short distances which is consistent with the equivalence principle. The creation and annihilation operators for the rest of $\alpha$ vacuua do not share this property. The annihlation operators for $\alpha$ vacua are related to the ones for the BD through Boguliobov transformations \cite{Bousso:2001mw},
\begin{align}\label{crea}
    a^\alpha_k=\frac{1}{\sqrt{1-e^{\alpha+\alpha^*}}}(a^{BD}_k-a_k^{BD~\dagger} e^{\alpha^*}).
\end{align}

The stability of de Sitter space is built in the symmetry group of it. So the instability of de Sitter space must manifest in the form of the impossibility of defining states that respect all de Sitter symmetries, i.e. de Sitter vacua. Following, we mention several issues with $\alpha$-vacua that are closely related to the instability of the de Sitter space. 
\begin{itemize}
    \item We use adiabatic approximation as we assume that for a comoving mode k, the state satisfying $a_k\ket\psi=0$ will continue satisfying it at later times. However, in \cite{Anderson:2017hts}, it was shown that this assumption does not hold. In other words, quantum effects make it impossible to find a true de Sitter vacuum that respects this symmetry. 
    
    \item Due to the previous point, the time at which the condition $a^{\alpha}_{k}\ket\psi=0$ is imposed for a given comoving mode matters. In \cite{Danielsson:2018qpa}, it was argued that the BD vacuum is the state we get by imposing these conditions at $t=-\infty$. This makes BD unphysical since, at that time, all the comoving modes are trans-Planckian. Physically, we can impose the annihilation conditions only after the modes enter the sub-Planckian regime. In \cite{Danielsson:2002kx}, this issue was resolved by introducing a more physical alternative to the BD vacuum by imposing the annihilation condition of each mode at the time it enters the sub-Planckian regime. This state is called the \textit{Instantaneous Minkowski Vacuum} (IMV). It was showed that this modification leads to decay of $\Lambda$ with a lifetime of $\sim H^{-1}$ \cite{Danielsson:2004xw,Danielsson:2005cc}. 

    \item Equation \eqref{crea} tells us that all $\alpha $-vacua (except BD) are UV divergent. This is because any $\alpha$-vacuum other than the BD vacuum has arbitrarily high momentum excitations with respect to the BD vacuum, which leads to a divergent energy-momentum tensor. The fact that the short distance behavior of $\alpha$-vacua are different from Minkowski violates the equivalence principle. 
\end{itemize}

The $\alpha$-vacua must be regulated at short distances to avoid the last problem. This could be done by introducing a cutoff $\Lambda$ and imposing $a_k^{BD}\ket{\alpha}_{reg}=0$ for $k>\Lambda$ and $a_k^{\alpha}\ket{\alpha}_{reg}=0$ for $k<\Lambda$. Note that the two-point function at scale $\sim 1/\Lambda$ is no longer invariant under de Sitter space isometries. This is another example that preserving the symmetries of de Sitter space at all scales is impossible.

The regulated $\alpha$-vacua $\ket{\alpha}_{reg}$ differ from BD only for modes with $k<\Lambda$. All of these modes exit the Hubble horizon in $\frac{1}{H}\ln(\frac{\Lambda}{H})$. In other words, after $\frac{1}{H}\ln(\frac{\Lambda}{H})$ any de Sitter vacuum evolves into the BD vacuum which is a thermal background. This result is identical to what we found in the previous subsection from de Sitter complementarity. 

\subsection{Complementarity in de Sitter space}\label{33}

Both of the perspectives that we discussed point towards the same expression $\frac{1}{H}\ln(\frac{1}{H})$ for de Sitter thermalization/scrambling time which matches the conjectured value in \cite{Susskind:2011ap}. The thought experiment in \ref{DSC1} gives us a unique insight into complementarity. It tells us that after the scrambling time, all the Hubble patch information exits the Hubble horizon and gets radiated back in the form of Hawking radiation. Let us say we have put observers on a comoving lattice with the initial spacing of Planck length. After a scrambling time, all observers exit each other's respective Hubble horizons. In a sense, each observer gets their own universe! Each observer will see all the other ones exit their horizon and receive their information in Hawking radiation after the scrambling time. There will be many isolated universes, each having a copy of the initial information in the form of Hawking radiation of everything else that crossed their Hubble horizon. This is the complementarity picture of de Sitter space. 

\section{Complementarity and the Swampland}\label{4}

In the last section, we provided arguments from different standpoints that the scrambling time in a de Sitter background is of the order of TCC time. The Swampland conditions suggest that the de Sitter space cannot be viewed as an equilibrium thermal background. This, however, does not mean that the de Sitter space does not have any statistical interpretation. For instance, the de Sitter entropy is still a meaningful quantity that counts the number of quasi-stable ds microstates. However, this should be viewed as a fine-grained entropy not to be confused with the thermodynamic entropy, which satisfies the second law of thermodynamics (see \cite{Almheiri:2020cfm} for a review of fine and coarse-grained entropies). For the thermodynamic entropy to make sense, the system must be able to reach equilibrium. In a sense, the complementarity picture breaks down according to TCC. 

It is worth mentioning that de Sitter space, if viewed as a thermal background, has some strange features. For example, the number of particles in the thermal radiation is $\mathcal{O}(1)$ \cite{Danielsson:2002td}. So in a sense, the de Sitter space would be the minimal thermodynamical system that quantum mechanically could make sense. 

There are some key differences between black hole complementarity and de Sitter complementarity. For example, Hawking-Page phase transition has no analogue in de Sitter space \cite{Parikh:2008iu}. Another fundamental difference between the two is that the horizon is real and covariant in the black hole version. In contrast, in de Sitter space, the horizon is apparent and observer dependent that could significantly differ from the real horizon. The difference between real and apparent horizons is significant especially for fastly decaying de Sitter spaces such as those predicted by TCC. It is intriguing to see if there is a modified version of the complementarity principle that applies to all real horizons and is consistent with the Swampland picture. 

A nice demonstration of the tension between Swampland conditions other than TCC and thermal aspects of de Sitter space can be found in \cite{Seo:2019wsh}. The number of light degrees of freedom in a 4d de Sitter space is given by the de Sitter entropy $N_\Lambda\sim \frac{1}{\Lambda}$. Applying the de Sitter conjecture and the distance conjecture to a rolling quintessence potential shows that the number of accessible degrees of freedom increases by more than $N_\Lambda$ over a scrambling time \cite{Seo:2019wsh}. Therefore, the low energy EFT breaks before the de Sitter space can thermalize any non-thermal perturbation.

\section{Cosmological implications}\label{5}

 The fluctuations of CMB are scale-invariant. The only $\alpha$-vacuum that generates scale-invariant fluctuations is the BD vacuum. This poses a fine-tuning problem for inflation's initial condition unless there is a natural mechanism that sets the vacuum to BD. As discussed in the last section, if the de Sitter space lasts more than the scrambling time, any initial vacuum would eventually evolve to BD. However, we saw that the Swampland conditions forbid this. Following, assuming a lifetime $\tau$ for the de Sitter space, we find the range of the $\alpha$ vacua that get turn into BD vacuum within the lifetime of de Sitter space. The smallness of this range determines the severity of the initial condition problem for inflation.
 
\begin{figure}
    \centering
    \includegraphics[scale=0.8]{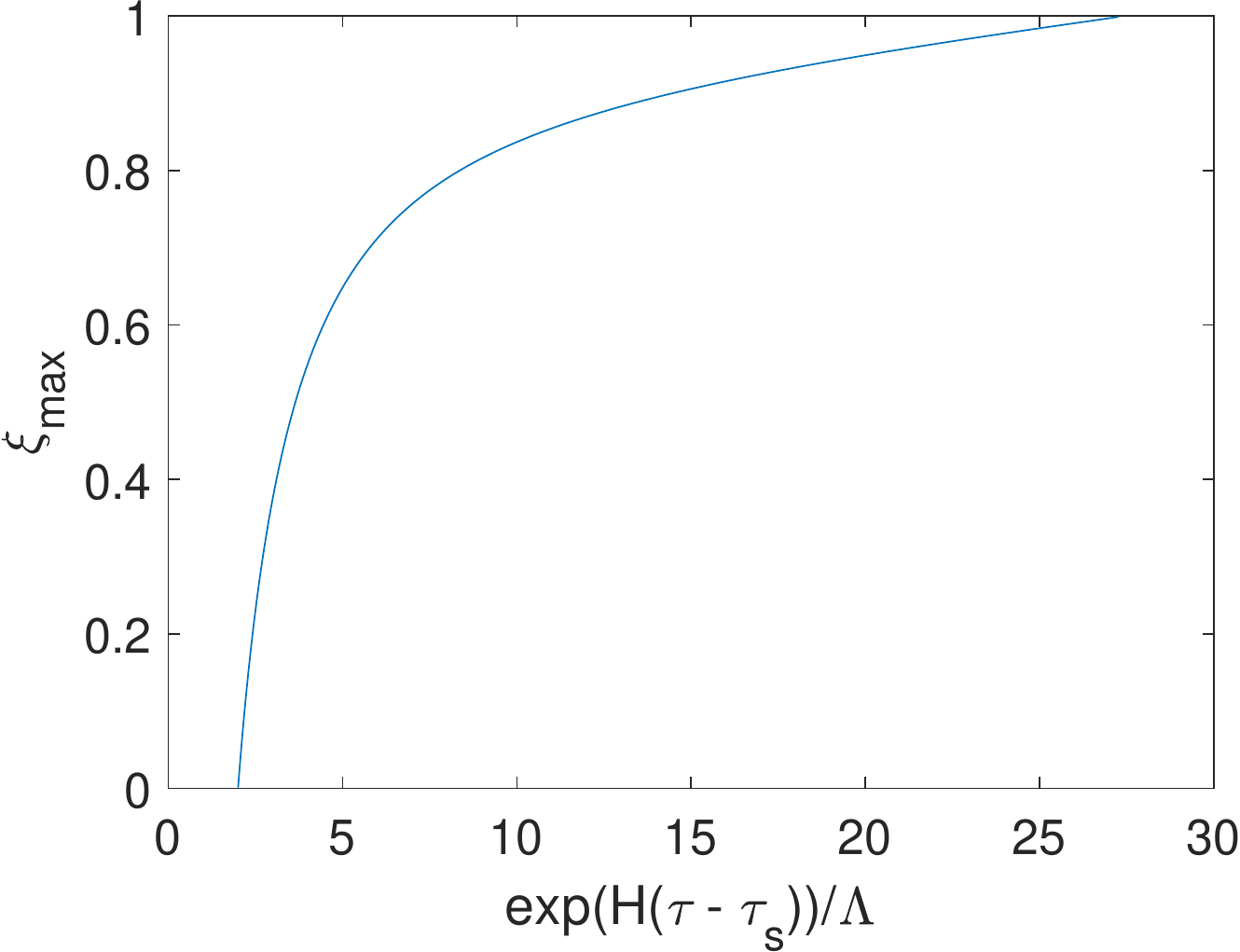}
    \caption{Plot of $\xi_{\max}$ versus $e^{H(\tau-\tau_s)}/\Lambda$.}
    \label{soft}
\end{figure}

Suppose $\ket\psi_k$ is the projection of $\ket \alpha$ over the Fock space of particles with momentum $k$. Let $\ket\psi_k=\sum_n c_n\ket n$, where $\ket n$ is the n-particle state with respect to BD modes. Since $a^\alpha\ket\psi_k=0$, from \eqref{crea} we find
\begin{align}
    c_{2i+1}=0~~and~~c_{2i}=\frac{(2i-1)!!}{\sqrt{(2i)!}}e^{i\alpha^*}.
\end{align}
Therefore, the average excitation number of the momentum-$k$ mode is
\begin{align}\label{excn}
    n_k\approx\frac{\sum_{i\geq0}2i\frac{(2i-1)!!}{\sqrt{(2i)!}}e^{2i\Re(\alpha)}}{\sum_{i\geq0}\frac{(2i-1)!!}{\sqrt{(2i)!}}e^{2i\Re(\alpha)}}.
\end{align}
For large $n$ we have
\begin{align}
    \ln\frac{(2n-1)!!}{\sqrt{(2n)!}}&= \frac{1}{2}\sum_{i=1}^n \ln(\frac{2i-1}{2i})\nonumber\\
    &=\frac{1}{2}\sum_{i=1}^n \ln(1-\frac{1}{2i})\nonumber\\
    &\approx \frac{1}{2}\sum_{i=1}^n -\frac{1}{2i}\nonumber\\
    &=-\frac{1}{4} H(n)\nonumber\\
    &\approx -\frac{1}{4} \ln(n),
\end{align}
where $H(n)$ is the harmonic series. Therefore, from exponentiating the above equation we find,
\begin{align}
    \frac{(2n-1)!!}{\sqrt{(2n)!}}\approx n^{-1/4}.
\end{align}
Plugging this into \eqref{excn} leads to
\begin{align}
        n_k\approx\frac{\sum_{i\geq0}2i^{3/4}e^{2i\Re(\alpha)}}{\sum_{i\geq0}i^{-1/4}e^{2i\Re(\alpha)}}.
\end{align}
This can be expressed in terms of the polylogarithm functions as
\begin{align}
    n_k\approx f(\xi),
\end{align}
where $\xi=\exp(2\Re(\alpha))$ measures the deviation from the Bunch-Davis vacuum and 
\begin{align}
    f(x):=2\frac{Li_{-\frac{3}{4}}(x)}{Li_{\frac{1}{4}}(x)}.
\end{align}
The average frequency is 
\begin{align}
    \expval{k}&\sim\frac{\int_{|k|<\Lambda} d^{D-1}k ~~ n_kk}{\int_{|k|<\Lambda} d^{D-1}k}\nonumber\\
    &\sim \Lambda f(\xi).
\end{align}
where $\Lambda$ is the field theory cut-off and $D$ is the dimension of space-time. The time that it takes for these deviations to freeze out (thermalize) is 
\begin{align}
    \tau_\xi=\frac{1}{H}\ln(\frac{\Lambda f(\xi)}{H}).
\end{align}
For the the regulated $\ket\alpha$ to thermalize into BD vacuum with scale invariant fluctuations, the lifetime of de Sitter space must be greater than this time
\begin{align}
    \tau>\frac{1}{H}\ln(\frac{\Lambda f(\xi)}{H}),
\end{align}
which can be rearranged into
\begin{align}
    \xi<f^{-1}(e^{H(\tau-\tau_s)}/\Lambda).
\end{align}
Figure \ref{soft} shows the graph of $\xi_{\max}=f^{-1}(e^{H(\tau-\tau_s)}/\Lambda)$.

In order to avoid an initial condition problem in slow-roll inflation, we should have $1-\xi_{\max}\ll1$. From figure \ref{soft} we can see that this would require $e^{H(\tau-\tau_s)}/\Lambda\gg 1$. If $\Lambda=O(M_{pl})$ we would need $\tau\gg\tau_s$ which is inconsistent with TCC. Thus, TCC implies that de Sitter space is not stable enough to allow all the quantum details to thermalize. In our universe, the fluctuations are very close to BD Gaussian fluctuations. This imposes a severe fine-tuning problem on inflation. This is in addition to another fine-tuning problem that TCC imposes on inflationary models due to the very short field range of the inflaton \cite{Brandenberger:2012aj}. In conclusion, any conventional form of inflation seems to be in severe tension with TCC. A TCC-compatible potential alternative for inflation was recently proposed by Prateek Agrawalet. al. for the early phase of our universe \cite{Agrawal:2020xek}.

\section{Conclusions}

We saw from two different points of view that suppose de Sitter space lives long enough, the time $\frac{1}{H}\ln(\frac{1}{H})$ could be viewed as the thermalization time. From the complementarity standpoint, this is when out-of-equilibrium perturbations thermalize over the stretched horizon before getting radiated back into the Hubble patch. From another point of view, this is when deviations from the thermal BD vacuum exit the Hubble horizon. These are two different, yet compatible, ways of viewing the thermalization process in de Sitter space. 

The TCC states that the lifetime of de Sitter space is less than the de Sitter thermalization time. In other words, the universe will quickly access more light degrees of freedom than the ones available to it in a give de Sitter background and will not stay in the de Sitter Hilbert space long enough to reach thermal equilibrium. Because of this, TCC poses a severe initial condition problem for any conventional inflationary scenario producing the scale-invariant CMB fluctuations. 

It would be interesting to study the possibility of a more general principle that quantum gravity forbids finite-dimensional thermal systems in the sense that the thermal distribution in any finite-dimensional subspace cannot be confined to that subspace for more than its thermalization time.  
\vspace{20pt}

\hspace{-15pt}\large\textbf{Acknowledgements}\normalsize
\vspace{10pt}

\hspace{-15pt}I am grateful to Cumrun Vafa for many insightful discussions and his generous support and guidance. I would like to thank Georges Obied, Matthew Reece, Mahdi Torabian, and Irene Valenzuela for helpful discussions and valuable comments. I would also like to thank Gary Shiu for valuable feedback on the manuscript.

This work was supported in part by the NSF grant PHY-2013858.

\appendix
\section{Thought experiment: Meeting beyond the Hubble horizon }\label{31}

We present the de Sitter version of the thought experiment discussed in subsection \ref{rbhc}. This thought experiment was developed in conversations with Cumrun Vafa and Georges Obied. We show that $\tau_s\gtrsim\tau_{TCC}$.

We assume that the vacuum is BD after a scrambling time. As we will see, this is consistent with our final result $\tau_s\gtrsim\tau_{TCC}$, because as we showed in subsection \ref{3p3}, any vacuum evolves into BD in $\tau_{TCC}$. Since BD vacuum is maximally entangled across the Hubble horizon, we can use it to perform the Hayden-Perskill protocol \cite{Hayden:2007cs} to recover the information. Therefore, the information of a system that has exited the Hubble horizon can be recovered after the scrambling time. 

Consider Alice and Bob carrying two fully entangled q-bits. The idea is to have Bob cross Alice's horizon and see if Alice can get two copies of Bob's state; one through Hawking radiation and another via a null signal from Bob. If Alice succeeds, both the no-cloning theorem and the monogamy theorem would be violated. 

Alice is initially stationary with respect to the comoving frame. Bob crosses Alice's stretched horizon located at $\sim l_{P}$ from the Hubble horizon and the spacetime point X. After he is one $l_{P}$ outside the Hubble horizon, he ,makes a measurement on the q-bit and sends the outcome by a null ray toward Alice. We consider an extra $l_{P}$ since any emergent phenomenon from quantum gravity such as a horizon has a Planckian resolution.

We call the spacetime point at which Bob sends the signal Y. As Bob jumps in, the information of the qbit he is carrying will thermalize and radiate back to Alice from the stretched horizon after a scrambling time $t_s\simeq 1/H\log(1/H)$. We denote the point of radiation by Z. Suppose Alice moves toward the horizon on a null ray and catches the signal midway at spacetime point T. 

We consider a de Sitter space with flat coordinates such that $t_X=0$ and the scale factor at $X$ is set to 1. 

The metric takes the form
\begin{align}
    ds^2=dt^2-a(t)^2[dr^2+d\Omega^2],
\end{align}
where $a(t)=e^{Ht}$. Therefore, $r_X=\frac{1}{H}-1$ and $t_X=0$. From $ds^2\geq0$ we find $t_Y-t_X\geq r_Y-r_X$.By plugging in $r_X=\frac{1}{H}-1$ and $r_Y=\frac{1}{H}+1$ we find $t_Y\geq2$. Bob's information radiate back off of the stretched horizon after a scrambling time $\tau_s$. Thus $t_Z\sim \tau_s$. The physical distance of $Z$ which is on the stretched horizon from the Hubble horizon is $\sim l_{P}$, therefore
\begin{align}
    a(t_Z)r_Z=\frac{1}{H}-1\rightarrow r_Z=e^{-H\tau_s}(\frac{1}{H}-1).
\end{align}
Now we solve the null ray equation to find when the radiation will reach Alice at T. 
\begin{align}
    &ds=0\rightarrow dt=-a(t)dr\nonumber\\
    &\rightarrow \Delta\frac{1}{H} e^{-Ht}=\Delta r\nonumber\\
    &\rightarrow e^{-Ht_T}-e^{-H\tau_s}=-H\frac{e^{-H\tau_s}(\frac{1}{H}-1)}{2}\nonumber\\
    &\rightarrow e^{-Ht_T}=e^{-H\tau_s}(H+\frac{1}{2}).
\end{align}
Now to see if there is any cloning paradox we should see if the future lightcones of T and Y intersect. For points in the future light cone of T, we have,
\begin{align}\label{eqts1}
    dt\geq a(t)dr&\rightarrow \frac{1}{H}(e^{-Ht_T}-e^{-Ht})\geq r\nonumber\\
    &\rightarrow r\leq \frac{1}{H}(e^{-H\tau_s}(H+\frac{1}{2})-e^{-Ht}).
\end{align}
For points in the future light cone of Y, we have,
\begin{align}\label{eqts2}
    dt\geq -a(t)dr&\rightarrow \frac{1}{H}(-e^{-Ht_Y}+e^{-Ht})\leq r-r_Y\nonumber\\
    &\rightarrow r\geq \frac{1}{H}+1-\frac{1}{H}(e^{-2H}-e^{-Ht})=\frac{e^{-Ht}}{H}+1+\frac{1-e^{-2H}}{H}.
\end{align}
From \eqref{eqts1} and \eqref{eqts2} we find the following inequality must hold to prevent the future lightcones of T and Y from intersecting so that the no-cloning theorem is not violated.
\begin{align}
    e^{-H\tau_s}(1+\frac{1}{2H})\leq 1+\frac{1-e^{-2H}}{H}.
\end{align}
For sub-planckian energy dinsities $H<1$, the LHS is $e^{-H\tau_s}\mathcal{O}(1)$ and the RHS is $\mathcal{O}(1)$. 
\begin{align}
     e^{-H\tau_s}\lesssim H\rightarrow \tau_s\gtrsim \frac{1}{H}\ln(\frac{1}{H}).
\end{align}

Note that if Bob could send his signal right after exiting the horizon instead of $l_{P}$ beyond the horizon, Alice could catch it and the experiment would fail. The Planckian resolution of the stretched horizon plays an important role in preventing a cloning paradox. 

\begin{figure}[H]
    \centering

\tikzset{every picture/.style={line width=0.75pt}} %set default line width to 0.75pt        

\begin{tikzpicture}[x=0.75pt,y=0.75pt,yscale=-1,xscale=1]
%uncomment if require: \path (0,551); %set diagram left start at 0, and has height of 551

%Shape: Rectangle [id:dp28250530412390806] 
\draw   (174.77,18.52) -- (469.93,18.52) -- (469.93,307.57) -- (174.77,307.57) -- cycle ;
%Straight Lines [id:da010384721675165354] 
\draw    (174.77,18.52) -- (469.93,307.57) ;
%Straight Lines [id:da2689154214119136] 
\draw [color={rgb, 255:red, 226; green, 95; blue, 93 }  ,draw opacity=1 ]   (190.43,46.47) -- (196.29,51.91) .. controls (198.64,51.82) and (199.87,52.96) .. (199.96,55.31) .. controls (200.05,57.66) and (201.27,58.8) .. (203.62,58.71) .. controls (205.97,58.62) and (207.2,59.76) .. (207.29,62.11) .. controls (207.38,64.46) and (208.6,65.6) .. (210.95,65.52) .. controls (213.3,65.43) and (214.53,66.57) .. (214.62,68.92) .. controls (214.71,71.27) and (215.93,72.41) .. (218.28,72.32) .. controls (220.63,72.23) and (221.86,73.37) .. (221.95,75.72) .. controls (222.04,78.07) and (223.27,79.21) .. (225.62,79.12) .. controls (227.97,79.03) and (229.19,80.17) .. (229.28,82.52) .. controls (229.37,84.87) and (230.6,86.01) .. (232.95,85.92) .. controls (235.3,85.83) and (236.52,86.97) .. (236.61,89.32) .. controls (236.7,91.67) and (237.93,92.81) .. (240.28,92.72) .. controls (242.63,92.63) and (243.85,93.77) .. (243.94,96.12) .. controls (244.03,98.47) and (245.26,99.61) .. (247.61,99.52) -- (247.97,99.86) -- (247.97,99.86) ;
\draw [shift={(188.23,44.43)}, rotate = 42.85] [fill={rgb, 255:red, 226; green, 95; blue, 93 }  ,fill opacity=1 ][line width=0.08]  [draw opacity=0] (10.72,-5.15) -- (0,0) -- (10.72,5.15) -- (7.12,0) -- cycle    ;
%Curve Lines [id:da22826486642683697] 
\draw [color={rgb, 255:red, 0; green, 0; blue, 0 }  ,draw opacity=1 ] [dash pattern={on 4.5pt off 4.5pt}]  (174.77,18.52) .. controls (199.56,45.22) and (286.9,138.4) .. (283.6,149.27) ;
%Curve Lines [id:da3839267627064198] 
\draw [color={rgb, 255:red, 255; green, 159; blue, 0 }  ,draw opacity=1 ][line width=0.75]    (277.46,140.99) .. controls (277.91,126.27) and (283.4,112.83) .. (292.02,101.93) ;
\draw [shift={(293.26,100.4)}, rotate = 489.88] [color={rgb, 255:red, 255; green, 159; blue, 0 }  ,draw opacity=1 ][line width=0.75]    (6.56,-1.97) .. controls (4.17,-0.84) and (1.99,-0.18) .. (0,0) .. controls (1.99,0.18) and (4.17,0.84) .. (6.56,1.97)   ;
%Shape: Star [id:dp8979385950140799] 
\draw  [color={rgb, 255:red, 0; green, 0; blue, 0 }  ,draw opacity=0 ][fill={rgb, 255:red, 236; green, 107; blue, 107 }  ,fill opacity=1 ] (249.21,97.31) -- (250.45,99.86) -- (253.22,100.26) -- (251.22,102.24) -- (251.69,105.04) -- (249.21,103.72) -- (246.73,105.04) -- (247.21,102.24) -- (245.2,100.26) -- (247.97,99.86) -- cycle ;
%Shape: Star [id:dp5537819928695482] 
\draw  [color={rgb, 255:red, 0; green, 0; blue, 0 }  ,draw opacity=0 ][fill={rgb, 255:red, 0; green, 0; blue, 0 }  ,fill opacity=1 ] (277.46,134.28) -- (278.7,136.94) -- (281.47,137.37) -- (279.46,139.45) -- (279.94,142.37) -- (277.46,140.99) -- (274.99,142.37) -- (275.46,139.45) -- (273.46,137.37) -- (276.22,136.94) -- cycle ;
%Straight Lines [id:da7751286842194862] 
\draw    (269.93,142.57) -- (243.93,108.57) ;
\draw [shift={(243.93,108.57)}, rotate = 412.59000000000003] [color={rgb, 255:red, 0; green, 0; blue, 0 }  ][line width=0.75]    (0,3.35) -- (0,-3.35)   ;
\draw [shift={(269.93,142.57)}, rotate = 412.59000000000003] [color={rgb, 255:red, 0; green, 0; blue, 0 }  ][line width=0.75]    (0,3.35) -- (0,-3.35)   ;
%Straight Lines [id:da10643326309563639] 
\draw [color={rgb, 255:red, 106; green, 93; blue, 226 }  ,draw opacity=1 ]   (213.04,20.09) -- (218.78,25.67) .. controls (221.13,25.64) and (222.33,26.8) .. (222.36,29.15) .. controls (222.39,31.51) and (223.59,32.67) .. (225.95,32.64) .. controls (228.3,32.61) and (229.5,33.77) .. (229.53,36.12) .. controls (229.56,38.48) and (230.76,39.64) .. (233.12,39.61) .. controls (235.47,39.58) and (236.67,40.74) .. (236.7,43.09) .. controls (236.73,45.45) and (237.93,46.61) .. (240.29,46.58) .. controls (242.65,46.54) and (243.85,47.7) .. (243.88,50.06) .. controls (243.91,52.42) and (245.1,53.58) .. (247.46,53.55) .. controls (249.82,53.51) and (251.02,54.67) .. (251.05,57.03) .. controls (251.08,59.39) and (252.27,60.55) .. (254.63,60.52) .. controls (256.99,60.48) and (258.19,61.64) .. (258.22,64) .. controls (258.25,66.36) and (259.44,67.52) .. (261.8,67.49) .. controls (264.16,67.45) and (265.36,68.61) .. (265.39,70.97) .. controls (265.42,73.33) and (266.61,74.49) .. (268.97,74.46) .. controls (271.33,74.42) and (272.53,75.58) .. (272.56,77.94) .. controls (272.59,80.3) and (273.78,81.46) .. (276.14,81.43) .. controls (278.5,81.39) and (279.7,82.55) .. (279.73,84.91) .. controls (279.76,87.27) and (280.95,88.43) .. (283.31,88.4) .. controls (285.67,88.36) and (286.87,89.52) .. (286.9,91.88) .. controls (286.93,94.24) and (288.12,95.4) .. (290.48,95.37) .. controls (292.84,95.34) and (294.04,96.5) .. (294.07,98.86) -- (295.87,100.61) -- (295.87,100.61) ;
\draw [shift={(210.89,18)}, rotate = 44.19] [fill={rgb, 255:red, 106; green, 93; blue, 226 }  ,fill opacity=1 ][line width=0.08]  [draw opacity=0] (10.72,-5.15) -- (0,0) -- (10.72,5.15) -- (7.12,0) -- cycle    ;
%Shape: Star [id:dp9082095798009056] 
\draw  [color={rgb, 255:red, 0; green, 0; blue, 0 }  ,draw opacity=0 ][fill={rgb, 255:red, 107; green, 135; blue, 236 }  ,fill opacity=1 ] (295.53,93.57) -- (296.83,96.28) -- (299.71,96.71) -- (297.62,98.82) -- (298.12,101.8) -- (295.53,100.39) -- (292.95,101.8) -- (293.44,98.82) -- (291.35,96.71) -- (294.24,96.28) -- cycle ;
%Straight Lines [id:da7590921184964339] 
\draw [color={rgb, 255:red, 65; green, 117; blue, 5 }  ,draw opacity=1 ][line width=1.5]    (174.54,57.86) -- (174.93,168.57) ;
%Straight Lines [id:da25686599187895043] 
\draw [color={rgb, 255:red, 65; green, 117; blue, 5 }  ,draw opacity=1 ][line width=1.5]    (210.42,18.52) -- (174.54,57.86) ;
%Shape: Star [id:dp8057119426865056] 
\draw  [color={rgb, 255:red, 0; green, 0; blue, 0 }  ,draw opacity=0 ][fill={rgb, 255:red, 0; green, 0; blue, 0 }  ,fill opacity=1 ] (187.37,39) -- (188.5,41.57) -- (191.05,41.98) -- (189.21,43.98) -- (189.64,46.8) -- (187.37,45.47) -- (185.09,46.8) -- (185.53,43.98) -- (183.69,41.98) -- (186.23,41.57) -- cycle ;
%Straight Lines [id:da08652347896437673] 
\draw    (174.77,307.57) -- (469.93,18.52) ;

% Text Node
\draw (218.67,38.53) node [anchor=north west][inner sep=0.75pt]  [rotate=-45.48] [align=left] {{\fontfamily{ptm}\selectfont {\footnotesize Hubble Horizon}}};
% Text Node
\draw (286.57,106.17) node [anchor=north west][inner sep=0.75pt]   [align=left] {{\fontfamily{ptm}\selectfont {\footnotesize Bob}}};
% Text Node
\draw (141.5,97.05) node [anchor=north west][inner sep=0.75pt]   [align=left] {{\fontfamily{ptm}\selectfont {\footnotesize Alice}}};
% Text Node
\draw (214.59,128.18) node [anchor=north west][inner sep=0.75pt]  [font=\scriptsize,rotate=-358.81]  {${\textstyle \Delta t\sim \tau _{s}}$};
% Text Node
\draw (273.07,150.72) node [anchor=north west][inner sep=0.75pt]   [align=left] {{\scriptsize {\fontfamily{ptm}\selectfont X}}};
% Text Node
\draw (299.53,83.12) node [anchor=north west][inner sep=0.75pt]   [align=left] {{\scriptsize {\fontfamily{ptm}\selectfont Y}}};
% Text Node
\draw (230.7,93.71) node [anchor=north west][inner sep=0.75pt]   [align=left] {{\scriptsize {\fontfamily{ptm}\selectfont Z}}};
% Text Node
\draw (182.93,56.65) node [anchor=north west][inner sep=0.75pt]   [align=left] {{\scriptsize {\fontfamily{ptm}\selectfont T}}};

\end{tikzpicture}

    \caption{Penrose diagram of the thought experiment in de Sitter space. Alice (green curve) and Bob (orange curve) carry entangled q-bits. The red and blue squiggly lines respectively represent the Hawking radiation and Bob's message both of which carry a copy of Bob's q-bit's information}
    \label{fig:my_label5}
\end{figure}
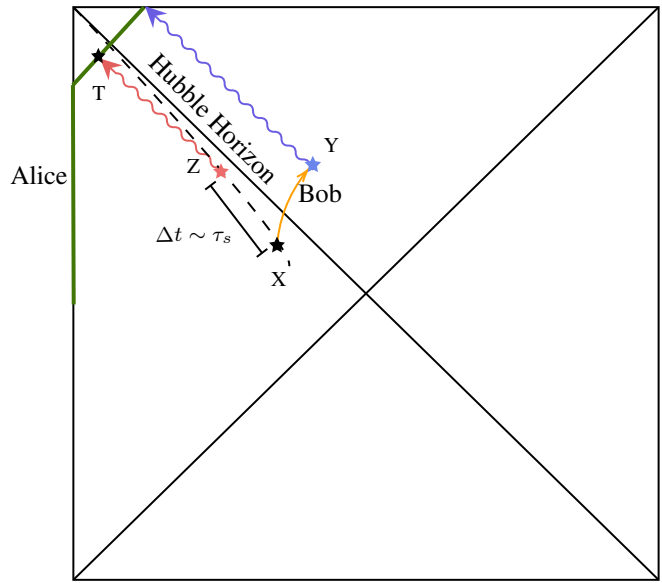

\bibliographystyle{unsrt}

\end{document}